\documentstyle[preprint,aps]{revtex}	
\begin{document}
%
%
\preprint{$
\begin{array}{l}
\mbox{UB--HET--97--03}\\[-3mm]
\mbox{UCD--97--20}\\[-3mm]
\mbox{October~1997}\\[5.mm]
\end{array}
$}
\title{QCD Corrections and Anomalous Couplings\\
in $Z\gamma$ Production at Hadron Colliders\\[5.mm]} 
\author{U.~Baur}
\address{
Department of Physics, State University of New York at Buffalo,
Buffalo, NY 14260, USA\\[3.mm]}
\author{T.~Han}
\address{
Department of Physics, University of California, Davis, CA 95616,
USA\\
{\rm and}\\
Department of Physics, University of Wisconsin, Madison, WI 53706,
USA\\[3.mm]}
\author{J.~Ohnemus}
\address{
Lawrence Berkeley Laboratory, Berkeley, CA 94720, USA\\[7.mm]}
\maketitle
\begin{abstract}
\baselineskip15.pt  
The processes $p\,p\hskip-7pt\hbox{$^{^{(\!-\!)}}$} \rightarrow 
Z \gamma + X \rightarrow \ell^+ \ell^- \gamma + X$ ($\ell=e,\,\mu$) and 
$p\,p\hskip-7pt\hbox{$^{^{(\!-\!)}}$} \rightarrow Z \gamma + X
\rightarrow \bar\nu\nu\gamma + X$ are calculated to ${\cal 
O}(\alpha_s)$ for general $ZZ\gamma$ and $Z\gamma\gamma$ couplings. 
The impact of ${\cal O}(\alpha_s)$ QCD corrections 
on the observability of $ZZ\gamma$ and $Z\gamma\gamma$ couplings
in $Z\gamma$ production at the Tevatron and the Large Hadron 
Collider (LHC) is discussed.
\end{abstract}
%
\newpage
%
%
\narrowtext

\section{INTRODUCTION}

Experiments at high energy hadron colliders provide an excellent opportunity 
to probe the interactions of the electroweak gauge bosons.
Recent observation of electroweak gauge boson pair production
at the Fermilab Tevatron have provided further confirmation of the
electroweak Standard Model (SM) and have significantly tightened 
the constraints on possible non-standard model self-interactions 
of the electroweak gauge 
bosons~\cite{cdfwgmdata,d0wgmdata,cdfwwdata,d0wwdata,d0dib}. 
For instance, $W\gamma$ production~\cite{cdfwgmdata,d0wgmdata}
can be used to study the $WW\gamma$ vertex, $WZ$ production can be used 
to probe the $WWZ$ vertex, and $W^+W^-$ 
production~\cite{cdfwwdata,d0wwdata,ell} is sensitive to both the 
$WW\gamma$ and $WWZ$ vertex functions. Furthermore, efforts have also 
been made to search for evidence of non-zero $ZZ\gamma$ and 
$Z\gamma\gamma$ couplings in $Z(\to\ell^+\ell^-)\gamma$ 
($\ell=e,\,\mu$)~\cite{d0dib,cdflldata,d0lldata} and 
$Z(\to\bar\nu\nu)\gamma$ production~\cite{d0dib,d0nndata}. 
These couplings vanish in the SM at tree level.
However, if new interactions beyond the SM are 
responsible for non-zero $ZZ\gamma$ or $Z\gamma\gamma$ couplings,
then $Z\gamma$ production could provide a clean signal for new physics.

Previous studies of $ZZ\gamma$ and $Z\gamma\gamma$ couplings in 
hadronic $Z\gamma$ production were based on leading order 
calculations~\cite{BAURBERGER}. Next-to-leading order calculations of 
$p\,p\hskip-7pt\hbox{$^{^{(\!-\!)}}$} \rightarrow Z\gamma$ production 
in the SM have shown that the NLO corrections are largest at high values of
the photon transverse momentum and high values of the $Z\gamma$ invariant
mass~\cite{NLOZGAMMA,NLOZGAMMADECAY}.  
These are the same regions of phase space where the effects
of non-standard $ZZ\gamma$ and $Z\gamma\gamma$ couplings 
are most pronounced~\cite{BAURBERGER}.
It is therefore important to include NLO corrections when probing
for evidence of non-standard $ZZ\gamma$ and $Z\gamma\gamma$ 
couplings in hadronic $Z\gamma$ production.
This paper presents a calculation of hadronic $Z(\to\ell^+\ell^-)
\gamma$ and $Z(\to\bar\nu\nu)\gamma$ production
to ${\cal O}(\alpha_s)$, including the most general non-standard
$ZZ\gamma$ and $Z\gamma\gamma$ couplings, and the decay of the
$Z$ boson in the narrow width approximation. Because of the larger 
$Z\to\bar\nu\nu$ branching ratio, the
$Z(\to\bar\nu\nu)\gamma$ cross section is about a factor~3 larger than
the combined $Z(\to e^+e^-)\gamma$ and $Z(\to \mu^+\mu^-)\gamma$ rates.
This results in substantially better limits on anomalous $ZZ\gamma$ and
$Z\gamma\gamma$ couplings in the $Z\to\bar\nu\nu$
channel~\cite{d0nndata}. 

To perform our calculation, we use the Monte Carlo method for NLO
calculations described in Ref.~\cite{NLOMC}. In the calculation the 
SM is assumed to be
valid apart from anomalies in the $ZZ\gamma$ and $Z\gamma\gamma$ vertices.
In particular, we assume the couplings of $W$ and $Z$ bosons to 
quarks and leptons to be given by the SM, and that there are no non-standard
couplings of the $Z\gamma$ pair to two gluons~\cite{BARBIERI}.
The gluon fusion process, $gg \to Z\gamma$, is small in the 
SM~\cite{GLUONFUSION} and is not considered in this paper.
Section~II briefly reviews the method used to carry out the calculation,
and describes how anomalous $ZZ\gamma$ and $Z\gamma\gamma$ couplings are
incorporated. 

In Sec.~III, we discuss how NLO QCD corrections influence the photon 
transverse momentum distribution, and derive sensitivity limits for 
non-standard $ZZ\gamma$ and $Z\gamma\gamma$ couplings at NLO for the 
Tevatron and LHC center-of-mass energies for various integrated
luminosities. The photon transverse momentum distribution is the
observable most sensitive to anomalous couplings~\cite{BAURBERGER}, and
is used by CDF and D\O\ to extract information on the $ZZ\gamma$ and
$Z\gamma\gamma$ vertices~\cite{d0dib,cdflldata,d0lldata,d0nndata}. 
At the Tevatron, QCD corrections are modest, and slightly improve the
sensitivity bounds. In contrast, at LHC energies, the inclusive NLO 
QCD corrections are quite large at high photon
transverse momenta in the SM and reduce the sensitivity to anomalous
$ZZ\gamma$ and $Z\gamma\gamma$ couplings somewhat. The large QCD
corrections are caused by a log squared enhancement factor in the $qg\to
Z\gamma q$ partonic cross section at high photon transverse momentum
($p_T$), and
the large quark-gluon luminosity at LHC energies. As in $W\gamma$~\cite{BHO}, 
$WZ$~\cite{BHO1,FRIX}, and $W^+W^-$ production~\cite{BHO2,WVENHANC}, the 
effect of the QCD corrections at high $p_T$ at the LHC can be reduced
by imposing a 0-jet requirement when searching for anomalous couplings. 
Finally, summary remarks are given in Sec.~IV.

\section{OVERVIEW OF THE CALCULATION}

The calculation presented here generalizes the results of 
Refs.~\cite{NLOZGAMMA} and~\cite{NLOZGAMMADECAY} to include general 
(non-standard model) $ZZ\gamma$ and $Z\gamma\gamma$ couplings.
The calculation employs a combination of analytic and Monte Carlo integration 
techniques; details of the method can be found in Ref.~\cite{NLOMC}.
The leptonic $Z$ boson decays are incorporated in the narrow width 
approximation. In this approximation,
radiative $Z$ decay diagrams, and graphs in which a virtual photon
decays into a charged lepton pair can be ignored. Radiative $Z$ decays are of
little interest when probing for non-standard couplings and can be
suppressed by a suitable choice of cuts (see Sec.~IIIB).
Furthermore, in the narrow width approximation it is particularly easy 
to extend the NLO calculation of hadronic $Z\gamma$ production in 
Ref.~\cite{NLOZGAMMA} to include the leptonic decay of the $Z$ boson.
The charged leptons are assumed to be massless in our calculation.

\subsection{Summary of ${\cal O}(\alpha_s)$ $Z\gamma$ production including
leptonic $Z$ decay}

\begin{figure}
\phantom{x}
\vskip 15cm
\includegraphics{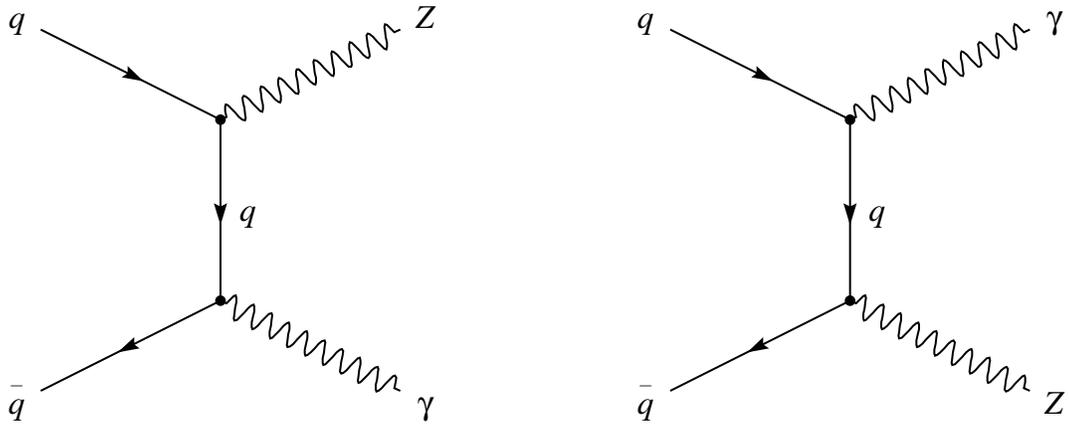}
\caption{Feynman diagrams for the Born level process 
$q \bar q \rightarrow Z \gamma$ in the Standard Model.}
\label{FIG:SMBORN}
\end{figure}

\begin{figure}
\phantom{x}
\vskip 15cm
\includegraphics{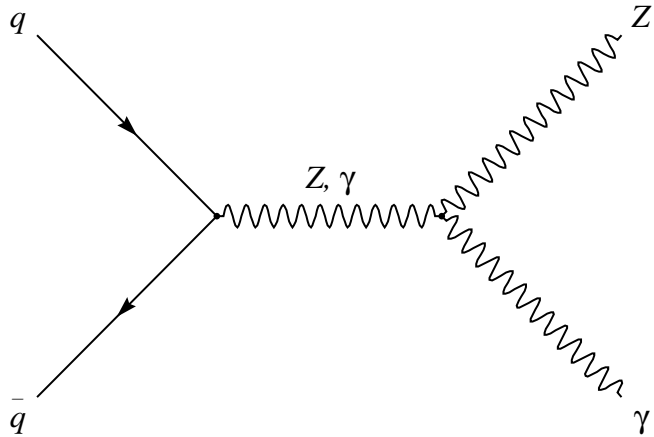}
\caption{Additional Feynman diagrams which contribute to the Born level 
process $q \bar q \rightarrow Z \gamma$ when non-standard model 
$ZZ\gamma$ and $Z\gamma\gamma$ couplings are introduced.}
\label{FIG:NONSMBORN}
\end{figure}

At lowest order in the SM, hadronic $Z\gamma$ production
proceeds via the Feynman diagrams shown in Fig.~\ref{FIG:SMBORN}.
Non-standard $ZZ\gamma$ and $Z\gamma\gamma$ couplings contribute via the
graphs shown in Fig.~\ref{FIG:NONSMBORN}. At the leading-logarithm level,
there are additional contributions to $Z\gamma$ production which come 
from photon bremsstrahlung processes such as $qg\to Zq$ followed by photon
bremsstrahlung from the final state quark. Although the process $qg\to
Zq$ is formally of ${\cal O}(\alpha\alpha_s)$, the photon fragmentation
functions are of order $\alpha/\alpha_s$~\cite{jeff}; thus the photon
bremsstrahlung process is of the same order as the Born process.

The NLO calculation of $Z\gamma$ production includes contributions from the
square of the Born graphs, the interference
between the Born graphs and the virtual one-loop diagrams, and the square 
of the real emission graphs. The basic idea of the method employed here 
is to isolate the
soft and collinear singularities associated with the real emission
subprocesses by partitioning phase space into soft, collinear, and
finite regions.  This is done by introducing theoretical soft and
collinear cutoff parameters, $\delta_s$ and $\delta_c$.  Using
dimensional regularization~\cite{DIMREG}, the soft and collinear
singularities are exposed as poles in  $\epsilon$ (the number of
space-time dimensions is $N = 4 - 2\epsilon$ with $\epsilon$ a small
number). The infrared singularities from the soft and virtual
contributions are then explicitly canceled while the collinear
singularities are factorized and absorbed into the definition of the
parton distribution functions. The remaining contributions are finite
and can be  evaluated in four dimensions.  The Monte Carlo program thus
generates $n$-body (for the Born and virtual contributions) and
$(n+1)$-body (for the real emission contributions) final state events. 
The $n$- and $(n+1)$-body contributions both depend on the cutoff
parameters $\delta_s$ and $\delta_c$, however, when these contributions
are added together to form a suitably inclusive observable, all
dependence on the cutoff parameters cancels.  
The numerical results presented in this paper are insensitive to 
variations of the cutoff parameters. 

Except for the virtual contribution, the ${\cal O}(\alpha_s)$ 
corrections are all proportional to the Born cross section. It is easy 
to incorporate the leptonic $Z$ decays into those terms which 
are proportional to the Born cross section; one simply replaces 
$d\hat\sigma^{\hbox{\scriptsize Born}} (q \bar q \to Z \gamma)$ with 
$d\hat\sigma^{\hbox{\scriptsize Born}} 
(q \bar q \to Z \gamma \to \ell^+ \ell^- \gamma)$ or 
$d\hat\sigma^{\hbox{\scriptsize Born}} 
(q \bar q \to Z \gamma \to \bar\nu\nu\gamma)$ in the relevant
formulae. When working at the amplitude level, the $Z$ boson decay
is trivial to implement; the $Z$ boson polarization vector,
$\epsilon_\mu (k)$, is simply replaced by the corresponding 
$Z \to \ell^+ \ell^-$ or $Z\to\bar\nu\nu$ decay current in the
amplitude.  Details of the amplitude level calculations for the Born
and real emission subprocesses can be found in Ref.~\cite{VVJET}.  

The only term in which it is more difficult to incorporate the $Z$
boson decay is the virtual contribution. Rather than undertake the 
non-trivial task of recalculating the virtual correction term for the 
case of a leptonically decaying $Z$ boson, we have instead opted to use 
the virtual correction for a real on-shell
$Z$ boson which we subsequently decay ignoring spin correlations.  
When spin correlations are ignored, the spin summed squared matrix element
factorizes into separate production and decay squared matrix elements.
Neglecting spin correlations slightly modifies the shapes of the 
angular distributions of the final state leptons, but does not alter
the total cross section as long as no angular cuts ({\it e.g.},
rapidity cuts) are imposed on the final state leptons. For realistic
rapidity cuts, cross sections are changed by typically 10\% when spin
correlations are neglected.
Since the size of the finite virtual correction is only about 
$2 - 4\%$ the size of the Born cross section, the overall effect of
neglecting the spin correlations in the finite virtual correction is expected
to be negligible compared to the combined $10 - 20\%$ uncertainty 
from the parton distribution functions, the choice of the factorization 
scale $Q^2$, and higher order QCD corrections. 

\subsection{General $ZZ\gamma$ and $Z\gamma\gamma$ Couplings}

\begin{figure}
\phantom{x}
\vskip 15cm
\includegraphics{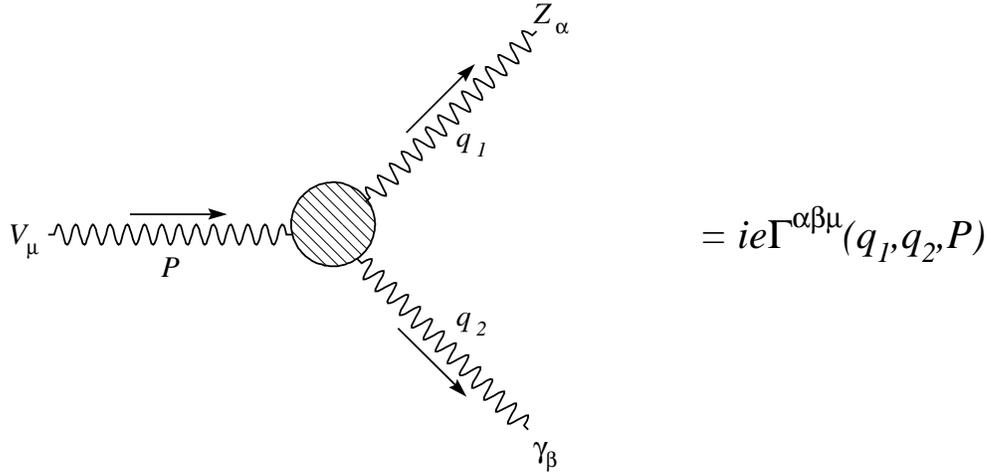}
\caption{Feynman rule for the general $Z\gamma V$ ($V= Z, \gamma$) vertex.
The factor $e$ is the charge of the proton.
The vertex function $\Gamma^{\alpha \beta \mu}_{Z\gamma V} (q_1, q_2, P)$ 
is given in Eq.~(\protect{\ref{EQ:GENVERTEX}}).}
\label{FIG:VERTEX}
\end{figure}

In $q \bar q \to Z\gamma$, the timelike virtual photon and/or 
$Z$ boson couples to essentially massless fermions, which ensures that
effectively $\partial_\mu V^\mu = 0$, $V=\gamma,\, Z$.  This fact, 
together with gauge
invariance of the on-shell photon, restricts the tensor structure of the
$Z\gamma V$ vertex sufficiently to allow only
four free parameters.  The most general non-standard $ZZ\gamma$
vertex function (see Fig.~\ref{FIG:VERTEX} for notation) 
is~\cite{LAGRANGIAN}
\begin{eqnarray}
\Gamma_{Z \gamma Z}^{\alpha \beta \mu} (q_1, q_2, P) =
{P^2 - q_1^2 \over m_Z^2} \Biggl\{ 
&\phantom{+}& h_1^Z \, \Bigl(  q_2^\mu    \, g^{\alpha \beta} 
                             - q_2^\alpha \, g^{\mu \beta}
                    \Bigr) \nonumber \\
&+& {h_2^Z \over m_Z^2} \, P^\alpha \, \Bigl( (P \cdot q_2) \, g^{\mu \beta}
                                       - q_2^\mu \, P^\beta 
                                 \Bigr) \nonumber \\
&+& h_3^Z \, \epsilon^{\mu \alpha \beta \rho} \, q_{2 \rho} 
\label{EQ:GENVERTEX} \\
&+& {h_4^Z \over m_Z^2} \, P^\alpha \, \epsilon^{\mu \beta \rho \sigma} \,
                        P_\rho \, q_{2 \sigma} 
\Biggr\} \>, \nonumber 
\end{eqnarray}
where $m_Z$ is the $Z$ boson mass.  The 
most general $Z\gamma\gamma$ vertex function can be obtained from 
Eq.~(\ref{EQ:GENVERTEX}) with the following replacements:
\begin{eqnarray}
{P^2 - q_1^2 \over m_Z^2} 
\longrightarrow
{P^2 \over m_Z^2} 
\qquad \hbox{\rm and} \qquad
h_i^Z \longrightarrow h_i^\gamma \>, \ i = 1, \ldots, 4 \>.
\label{EQ:REPLACEMENT}
\end{eqnarray}
Terms proportional to $P^\mu$ and $q_1^\alpha$ have been omitted in 
Eq.~(\ref{EQ:GENVERTEX}) since they do not contribute to the cross section.
Without loss of generality, the overall $ZZ\gamma$ and $Z\gamma\gamma$ 
coupling has been chosen to be
\begin{equation}
g_{ZZ\gamma}^{} = g_{Z\gamma\gamma}^{} = e \>,
\label{EQ:OVERALLCOUPLING}
\end{equation}
where $e$ is the charge of the proton.
The overall factor $(P^2 - q_1^2)$ in Eq.~(\ref{EQ:GENVERTEX})
is a result of Bose symmetry, whereas the factor $P^2$ in the $Z\gamma\gamma$
vertex function originates from electromagnetic gauge invariance.
As a result, the $Z\gamma\gamma$ vertex function vanishes identically if
both photons are on-shell due to Yang's theorem~\cite{YANG}. 

All the anomalous couplings $h_i^V$ ($i=1,\dots ,4$, $V=\gamma,\,Z$) 
are $C$ odd; $h_1^V$ and $h_2^V$ violate $CP$. $h_2^V$ and $h_4^V$ 
receive contributions only from operators of dimension $\geq 8$. Within 
the standard model, at tree level, all the couplings $h_i^V$ vanish.
At the one loop level, only the $CP$ conserving couplings 
$h_3^V$ and $h_4^V$ are nonzero~\cite{BARROSO}.
 
For simplicity, the $Z$ boson mass $m_Z$ has been chosen 
in Eq.~(\ref{EQ:GENVERTEX}) as the energy scale in the denominator of the
overall factor and the terms proportional to $h^V_{2,4}$.
For a different mass scale, $M$, all subsequent results can be obtained by
scaling $h^V_{1,3}$ $(h^V_{2,4})$ by a factor $M^2/m_Z^2$ $(M^4 / m_Z^4)$.

Tree level unitarity restricts the $ZZ\gamma$ and $Z\gamma\gamma$ couplings
uniquely to their standard model values at asymptotically high 
energies~\cite{CORNWALL}.
This implies that the $Z\gamma V$ couplings $h_i^V$ have to be described
by form factors $h_i^V (q_1^2, q_2^2, P^2)$ which vanish when
$q_1^2$, $q_2^2$, or $P^2$ becomes large.
In $Z\gamma$ production $q_2^2 = 0$ and $q_1^2 \approx m_Z^2$ even when 
finite $Z$ width effects are taken into account.  However, large values
of $P^2 = \hat s$ will be probed in future hadron collider experiments and
the $\hat s$ dependence has to be included in order to avoid unphysical
results that would violate unitarity. A detailed discussion of
unitarity and form factors in nonstandard $Z\gamma$ production 
can be found in Ref.~\cite{BAURBERGER}.
We will use generalized dipole form factors of the form
\begin{equation}
h_i^V (m_Z^2, 0, \hat s) = 
{ h_{i0}^V \over (1+\hat s/ \Lambda_{FF}^2 )^n } \>, 
\label{EQ:DIPOLES}
\end{equation}
as advocated in Ref.~\cite{BAURBERGER}.
The subscript $0$ denotes the low energy value of the form factor.
The mass scale $\Lambda_{FF}$ is the scale at which novel interactions,
like multiple weak boson or resonance production, may appear.
Unless stated otherwise, we use $n=3$ ($n=4$) for $h_{1,3}^V$ ($h_{2,
4}^V$) and $\Lambda_{FF}=500$~GeV [$\Lambda_{FF}=3$~TeV] at the Tevatron
[LHC].

At present, the most stringent direct bounds on anomalous $Z\gamma V$
couplings come from $Z\gamma$ production at the Tevatron collider,
$Z\to\bar\nu\nu\gamma$ decays at LEP1, and $\gamma+$~invisible particles
production at LEP2. From
a search performed in the channels $p\bar p\to Z(\to\ell^+\ell^-)
\gamma$ and $p\bar p\to Z(\to\bar\nu\nu)\gamma$,
the D\O\ Collaboration obtains from Run~1a data~\cite{d0dib}
\begin{equation}
|h_{30}^Z|<0.78~~({\rm for}~h_{40}^Z=0),\qquad\qquad |h_{40}^Z|<0.
19~~({\rm for}~h_{30}^Z=0),
\end{equation}
and
\begin{equation}
|h_{30}^\gamma|<0.81~~({\rm for}~h_{40}^\gamma=0),\qquad\qquad
|h_{40}^\gamma|<0.20~~({\rm for}~h_{30}^\gamma=0),
\end{equation}
at the 95\%~confidence level (CL). The limits obtained for $h_{10}^V$ 
and $h_{20}^V$
virtually coincide with those found for $h_{30}^V$ and $h_{40}^V$,
respectively. The L3 Collaboration obtains a slightly better bound 
on $h_{30}^Z$~\cite{L3}. From $e^+e^-\to\gamma+$~invisible particles at 
LEP2, the DELPHI collaboration finds a preliminary limit of
$|h_{30}^\gamma|<0.5$~\cite{DELPHI}. To derive these limits, the
experiments assumed a form factor scale of $\Lambda_{FF}=500$~GeV with 
$n=3$ ($n=4$) for $h_3^V$ ($h_4^V$). 

It is straightforward to include the non-standard model couplings in
the amplitude level calculations. We computed the $q \bar q \to Z \gamma$ 
virtual correction with the vertex function of Eq.~(\ref{EQ:GENVERTEX}) 
in the 't~Hooft-Veltman scheme~\cite{DIMREG,HV} using the computer
algebra program FORM~\cite{FORM}. The resulting expression, however, is 
too lengthy to present here.

Note that the non-standard $ZZ\gamma$ and $Z\gamma\gamma$ couplings do not
destroy the renormalizability of QCD. Thus the infrared singularities
from the soft and virtual contributions are explicitly canceled, and the
collinear singularities are factorized and absorbed into the definition
of the parton distribution functions, exactly as in the standard model case. 

The squared matrix element is bi-linear in the anomalous couplings. Due
to the antisymmetry of $\epsilon^{\mu \alpha \beta \rho}$, all terms
proportional to $h_3^V$, $h_4^V$, $h_3^V\,h_i^V$, and $h_4^V\,h_i^V$
($i=1,2$) vanish in the LO as well as in the NLO squared matrix
elements, after the sum over the fermion helicities and the photon
polarizations is performed. Terms
proportional to $h_1^V$ and $h_2^V$ are present in the squared matrix
elements. These terms are proportional to $\cos\theta^*$, where
$\theta^*$ is the scattering angle of the photon in the parton center of
mass frame. They vanish after the integration over phase space 
is performed, unless rapidity cuts on the photon which are not symmetric
with respect to $\eta=0$ (where $\eta$ is the pseudorapidity) are chosen.

\section{PHENOMENOLOGICAL RESULTS}

We now discuss the phenomenological implications of ${\cal 
O}(\alpha_s)$ QCD corrections and general $ZZ\gamma$ and 
$Z\gamma\gamma$ couplings in $Z\gamma$ production at the Tevatron
($p\bar p$ collisions at $\sqrt{s} = 1.8$~TeV) and the LHC ($pp$
collisions at $\sqrt{s} = 14$~TeV). First, 
the input parameters, cuts, and the finite energy resolution smearing
used to simulate detector response are briefly described.  We then
discuss in detail the impact of ${\cal O}(\alpha_s)$ QCD corrections 
on the observability of non-standard $ZZ\gamma$ and $Z\gamma\gamma$
couplings in $Z(\to\ell^+\ell^-)\gamma$ and $Z(\to\bar\nu\nu)\gamma$ 
production at the Tevatron and LHC. As mentioned in the Introduction, we
make no attempt to include the contributions from gluon fusion, $gg\to
Z\gamma$, which are formally of ${\cal O}(\alpha_s^2)$, into our
calculation. Gluon fusion contributes less than 0.2\% (6\%) to the total
$Z\gamma$ cross section at the Tevatron (LHC)~\cite{GLUONFUSION}.

\subsection{Input Parameters}

The numerical results presented here were obtained using the two-loop 
expression for $\alpha_s$. The QCD scale $\Lambda_{\hbox{\scriptsize QCD}}$ 
is specified for four
flavors of quarks by the choice of the parton distribution functions  and
is adjusted whenever a heavy quark threshold is crossed so that
$\alpha_s$ is a continuous function of $Q^2$. The heavy quark masses
were taken to be $m_b=5$~GeV and $m_t=176$~GeV~\cite{TOPMASS1,TOPMASS2}.

The SM parameters used in the numerical simulations are $m_Z = 91.187$~GeV,
$m_W = 80.22$~GeV, $\alpha (m_W) =1/128$, and $\sin^2
\theta_{\hbox{\scriptsize W}} = 1 - (m_W^{}/m_Z^{})^2$. These values are
consistent with recent measurements at LEP, SLC, the CERN $p\bar p$
collider, and the Tevatron~\cite{Marcel}. The soft and collinear 
cutoff parameters, discussed in Sec.~IIA, 
are fixed to $\delta_s = 10^{-2}$ and $\delta_c = 10^{-3}$. The parton
subprocesses have been summed over $u,d,s$, and $c$ quarks.
The $Z$ boson leptonic branching ratio is taken to be 
$B(Z \to e^+ e^-) = 0.034$ and the total width of the $Z$ boson is
$\Gamma_Z = 2.490$~GeV. Except where otherwise stated, a single scale
$Q^2=M^2_{Z\gamma}$,  where $M_{Z\gamma}$ is the invariant mass of the
$Z\gamma$ pair, has been used for the renormalization scale $\mu^2$
and the factorization scale $M^2$. 
The NLO numerical results have been calculated in the modified Minimal
Subtraction ($\overline {\rm MS}$) scheme~\cite{MSBAR}.

In order to get consistent NLO results it is necessary to use parton
distribution functions which have been fit to next-to-leading order. 
The numerical simulations have been performed using 
the Martin-Roberts-Stirling (MRS)~\cite{MRSA} set~A distributions  
($\Lambda_4 = 230$~MeV) 
in the $\overline {\rm MS}$ scheme.  For convenience,
the MRS set~A distributions have also been used for the LO calculations.

\subsection{Acceptance Cuts}

The cuts imposed in our numerical simulations are motivated by two factors:
1)~the finite acceptance and resolution of the detector, and
2)~the need to suppress radiative $Z$ decays which result in the same
final state as $Z\gamma$ production but which are of little interest for the
study of anomalous couplings in hadronic collisions. The finite
acceptance of the detector is simulated by cuts on the four-vectors of the
final state particles.  This group of cuts includes requirements on the 
transverse momentum of the photon and charged leptons for
$Z(\to\ell^+\ell^-)\gamma$, and on the missing transverse momentum, 
$p\llap/_T^{}$, resulting from the non-observation of the neutrinos in
$Z(\to\bar\nu\nu)\gamma$.
Also included in this group are cuts on the pseudorapidity, $\eta$, of the
photon and the charged leptons. In addition, the charged leptons and the
photon are
also required to be separated in the pseudorapidity-azimuthal-angle plane
\begin{eqnarray}
\noalign{\vskip 5 pt}
\Delta R (\ell,\gamma) = \left[ \left(\Delta \phi_{\ell \gamma}^{} \right)^2
	                      + \left(\Delta \eta_{\ell \gamma}^{} \right)^2
\right]^{1/2} \>.
\end{eqnarray}

Since we ignore photon radiation from the final state lepton line in our
calculation, it is necessary to impose cuts which will efficiently suppress 
contributions from these diagrams.
In radiative $Z$ decays the lepton-photon separation sharply peaks at
small values due to the collinear singularity associated with the
diagrams in which the photon is radiated from the final state lepton
line.  In the following we shall therefore impose a large separation cut
of $\Delta R(\ell,\gamma)>0.7$. Contributions from $Z\to \ell^+\ell^-\gamma$
can be further reduced by an invariant mass cut on the $\ell\ell\gamma$
system of $M(\ell\ell\gamma)>100$~GeV.

At leading order, $Z\gamma$ events are produced not only by the Born 
subprocess $q \bar q \rightarrow Z \gamma$ but also by the photon 
bremsstrahlung process which proceeds via subprocesses such as $qg 
\rightarrow Zq$ followed by photon bremsstrahlung from the final state 
quark. As demonstrated in Ref.~\cite{BREM}, the bremsstrahlung process is 
significant at LHC energies. However, this process does 
not involve the $Z\gamma V$ vertices and is thus a background in a 
search for anomalous $ZZ\gamma$ and $Z\gamma\gamma$ couplings.
Fortunately, the photon bremsstrahlung events can be suppressed by requiring
the photon to be isolated~\cite{BREM}. A photon isolation cut typically 
requires the sum of the hadronic energy $E_{\hbox{\scriptsize had}}$ in a
cone of size $R_0$ about the direction of the photon to be less than a
fraction $\epsilon_{\hbox{\scriptsize h}}$ of 
the photon energy $E_{\gamma}$, {\it i.e.},
\begin{eqnarray}
\sum_{\Delta R < R_0} \, E_{\hbox{\scriptsize had}} < 
\epsilon_{\hbox{\scriptsize h}} \, E_{\gamma} \>,
\label{EQ:ISOL}
\end{eqnarray}
with $\Delta R = [(\Delta \phi)^2 + (\Delta \eta)^2 ]^{1/2}$. 
To suppress the photon bremsstrahlung background, a photon isolation cut
with $\epsilon_{\hbox{\scriptsize h}}=0.15$ and $R_0=0.7$~\cite{EPS} will
be applied in the numerical results presented in this section. For this
value of $\epsilon_{\hbox{\scriptsize h}}$, the photon bremsstrahlung
background is less than a few per cent of the Born $Z\gamma$ signal rate.
The complete set of cuts for $Z(\to\ell^+\ell^-)\gamma$
is summarized in the following table.
\begin{quasitable}
\begin{tabular}{ccc}
\multicolumn{3}{c}{$Z\gamma \to \ell^+ \ell^- \gamma$ } \\
Tevatron & & LHC\\
\tableline
$p_{T}^{}(\gamma)      > 10$~GeV  & & $p_{T}^{}(\gamma)      > 100$~GeV\\
$p_{T}^{}(\ell)        > 15$~GeV  & & $p_{T}^{}(\ell)        > 20$~GeV\\
$|\eta(\gamma)|        < 3.0$     & & $|\eta(\gamma)|        < 3.0$\\
$|\eta(\ell)|          < 3.5$     & & $|\eta(\ell)|          < 3.0$\\
$\Delta R(\ell,\gamma) > 0.7$     & & $\Delta R(\ell,\gamma) > 0.7$\\
$M(\ell\ell\gamma)     > 100$~GeV & & $M(\ell\ell\gamma)     > 100$~GeV\\
$\sum_{\Delta R < 0.7} \, E_{\hbox{\scriptsize had}} < 0.15 \,
E_{\gamma}$ & & $\sum_{\Delta R < 0.7} \, E_{\hbox{\scriptsize had}} < 
0.15 \, E_{\gamma}$ 
\end{tabular}
\end{quasitable}
Since our calculation is carried out in the narrow width approximation
for the $Z$ boson, no explicit cut on the di-lepton invariant mass is imposed.

If the $Z$ boson decays into a pair
of neutrinos, the experimental signal is $p\,
p\hskip-7pt\hbox{$^{^{(\!-\!)}}$} \to\gamma p\llap/_T$, with the missing
transverse momentum, $p\llap/_T$, resulting from the nonobservation of the
neutrino pair. 
For $Z \gamma \to p\llap/_T^{} \gamma$ at the Tevatron we use the same
transverse momentum and pseudo-rapidity cuts as the D\O\ Collaboration
in their $Z(\to\bar\nu\nu)\gamma$ analysis~\cite{d0nndata}. The following
table summarizes the cuts imposed for both the Tevatron and LHC
analysis.
\begin{quasitable}
\begin{tabular}{cc}
\multicolumn{2}{c} {$Z \gamma \to p\llap/_T^{} \gamma$} \\
Tevatron  & LHC\\
\tableline
$p_{T}^{}(\gamma)      > 40$~GeV  &  $p_{T}^{}(\gamma)      > 100$~GeV\\
$p\llap/_{T}^{}        > 40$~GeV  &  $p\llap/_{T}^{}        >
100$~GeV\\
no jet with $p_T(j)>15$~GeV and $|\eta(j)|<2.5$ & 
no jet with $p_T(j)>50$~GeV  and $|\eta(j)|<3$ \\
$|\eta(\gamma)|        < 2.5$    &  $|\eta(\gamma)|        < 3.0$\\
\end{tabular}
\end{quasitable}
The high $p_T(\gamma)$ and $p\llap/_T$ cuts, when combined with the jet
veto, strongly reduce the background from $W\to e\nu$ events where the 
electron is misidentified as a photon at the Tevatron. These cuts
also eliminate backgrounds from $\gamma j$ production with the jet
rapidity outside the range covered by the detector and thus ``faking''
missing transverse momentum, and $jj$ production where one of the jets
is misidentified as a photon, while the other disappears through the
beam hole. The large
$p_T^{}(\gamma)$ and $p\llap/_T^{}$ cuts at LHC energies are chosen to
reduce potentially dangerous backgrounds from 
$Z+1$~jet production, where the jet is misidentified as a photon, and
from processes where particles outside the rapidity range covered by the
detector contribute to the missing transverse momentum. Present 
studies~\cite{CMS,ATLAS} indicate that these backgrounds are under control for 
$p_T^{}(\gamma) > 100$~GeV and $p\llap/_T^{}>100$~GeV. 

\subsection{Finite Energy Resolution Effects}

Uncertainties in the energy measurements of the charged leptons and photon 
in the detector are simulated in the calculation by Gaussian smearing 
of the particle four-momentum vector with standard deviation $\sigma$.
For distributions which require a jet definition, {\it e.g.}, the 
$Z\gamma + 1$~jet exclusive cross section, the jet four-momentum vector 
is also smeared. The standard deviation $\sigma$
depends on the particle type and the detector. The numerical results 
presented here for the Tevatron and LHC center of mass energies 
were obtained using $\sigma$ values based on the CDF~\cite{RCDF} and
ATLAS~\cite{ATLAS} specifications, respectively. For 
$Z(\to\bar\nu\nu)\gamma$ production at the Tevatron, the photon and
$p\llap/_T$ vectors were smeared using the D\O\ resolutions given in
Ref.~\cite{d0nndata}. 

\subsection{${\cal O} (\alpha_s)$ Corrections and Anomalous $ZZ\gamma$ 
and $Z\gamma\gamma$ Couplings in $Z(\to\ell^+\ell^-)\gamma$ Production}

Non-standard $ZZ\gamma$ and $Z\gamma\gamma$ couplings have a significant
effect on many distributions in $Z\gamma$ production. The photon 
transverse momentum distribution was found~\cite{BAURBERGER} to be the
distribution most sensitive to anomalous couplings.  We will therefore 
concentrate on it in the following presentation.

\begin{figure}
\phantom{x}
\vskip 15cm
\includegraphics{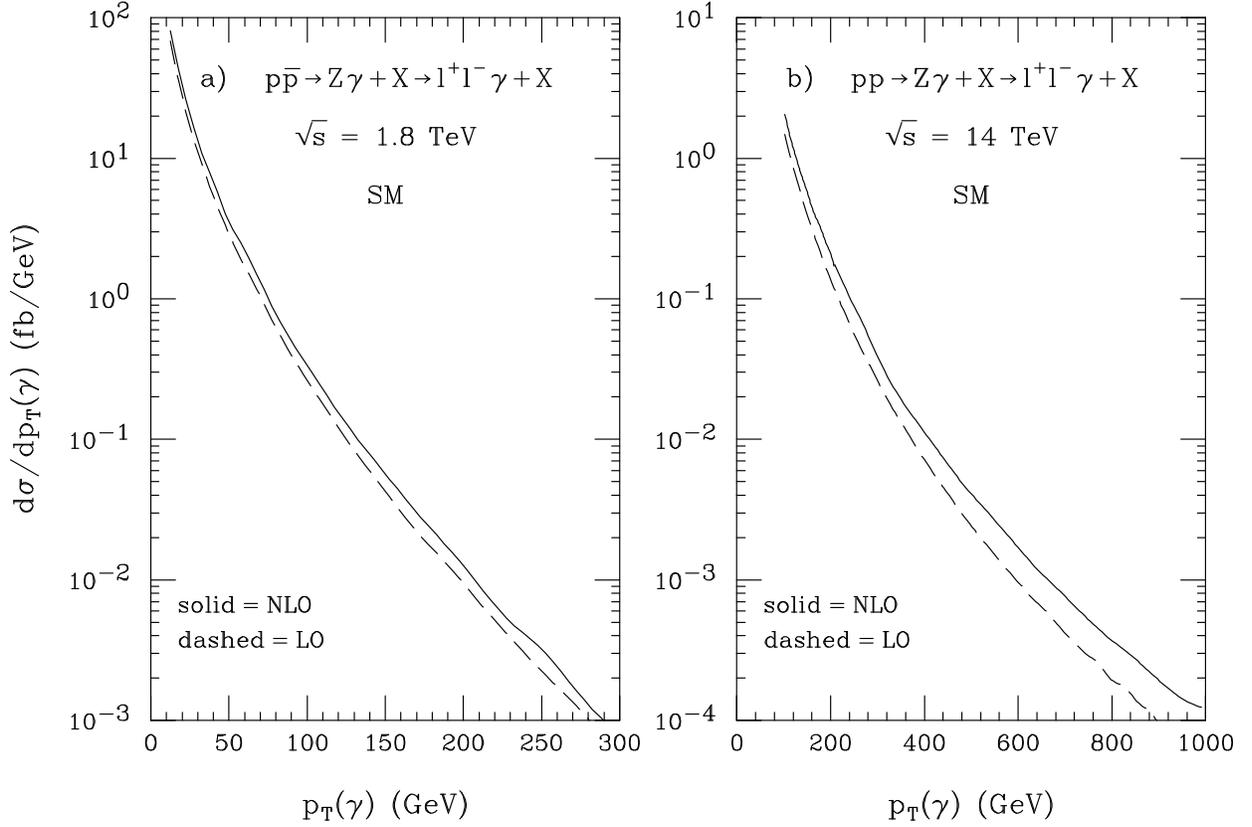}
\caption{Differential cross sections versus $p_T^{}(\gamma)$ for 
(a) $p\bar p \to Z\gamma +X \to \ell^+ \ell^- \gamma + X$ at
$\protect{\sqrt{s}=1.8}$~TeV, 
and (b) $p p \to Z\gamma +X \to \ell^+ \ell^- \gamma + X$ at
$\protect{\sqrt{s}=14}$~TeV in the SM.
The jet-inclusive cross sections are shown at the Born level 
(dashed curves) and with the NLO corrections (solid curves). The cuts
imposed are summarized in Sec.~IIIB.}
\label{FIG:NLO}
\end{figure}

The LO and NLO SM photon transverse momentum distributions for
$p\,p\hskip-7pt\hbox{$^{^{(\!-\!)}}$} \rightarrow Z\gamma + 
X\to\ell^+\ell^-\gamma + X$ production at Tevatron and LHC center of 
mass energies are shown in Fig.~\ref{FIG:NLO}. Here, and in all
subsequent results shown for $Z\to\ell^+\ell^-$ decays, we sum over
electron and muon final states. 
The NLO corrections grow with the photon transverse momentum
and with the center of mass energy. At the LHC, for example, the 
QCD corrections
increase the SM cross section by about a factor~2.2 at $p_T(\gamma)
=1$~TeV, whereas the enhancement is only a factor~1.4 at $p_T(\gamma)
=100$~GeV. The large QCD corrections at high values of $p_T(\gamma)$ are
caused by a collinear enhancement factor, $\log^2(p_T(\gamma)/m_Z)$, in the 
$qg\to Z\gamma q$ partonic cross
section for photon transverse momenta much larger than $m_Z$, 
$p_T(\gamma)\gg m_Z$, and the large $qg$ luminosity at LHC energies. 
The large corrections arise from the kinematical region where the photon is
produced at large $p_T^{}$ and recoils against the quark, which radiates a
soft $Z$ boson which is almost collinear to the quark, and thus is
similar in nature to the enhancement of QCD corrections observed at
large vector boson transverse momenta in $W\gamma$, $WZ$, and $W^+W^-$
production~\cite{BHO,FRIX,WVENHANC}. The effect, however, is less
pronounced than in these processes. In $W\gamma$ and $WZ$ production,
the SM Born cross section is suppressed due to the appearance of an exact
or approximate radiation zero~\cite{RADZERO,BHOR}, while there is no
radiation zero in the $Z\gamma$ case. In $W^+W^-$ production, the strong
correlation of the $W$ helicities in the SM, together with the effect of
kinematic cuts, is responsible for the larger effect of the QCD
corrections~\cite{BHO2}.

\begin{figure}
\phantom{x}
\vskip 15cm
\includegraphics{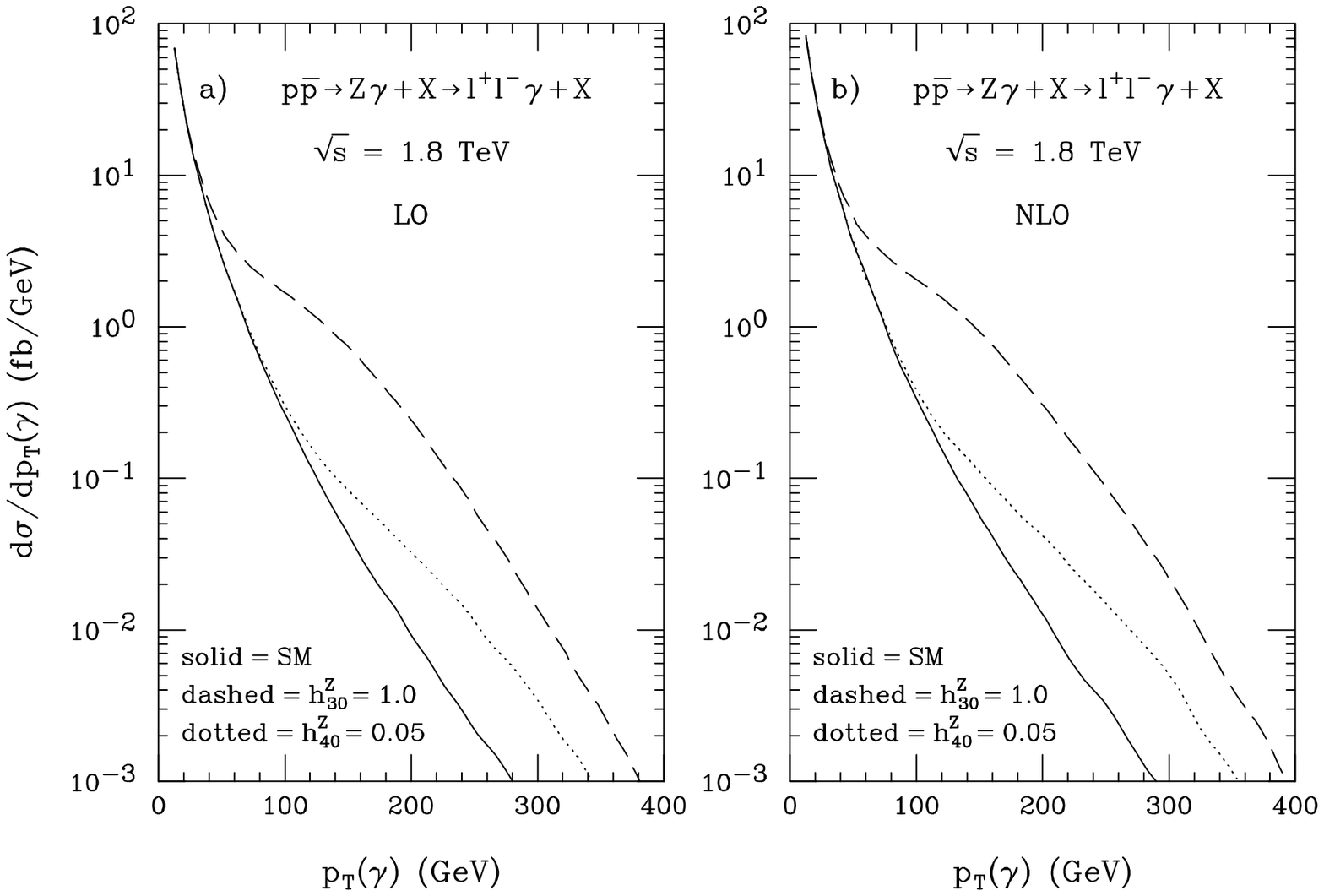}
\caption{The differential cross section for the photon transverse
momentum in the reaction $p\bar p \to Z\gamma +X \to \ell^+ \ell^- 
\gamma + X$ at $\protect{\sqrt{s}=1.8}$~TeV,
(a) in the Born approximation and (b) including NLO QCD corrections. 
The curves are for the SM (solid), $h_{30}^Z=1.0$ (dashed), and 
$h_{40}^Z=0.05$ (dotted). The cuts imposed are summarized in Sec.~IIIB.}
\label{FIG:ACTEV}
\end{figure}

The effects of non-standard $ZZ\gamma$ couplings on the photon 
transverse momentum distribution in $p\bar p\to Z\gamma + 
X\to\ell^+\ell^-\gamma + X$ at the Tevatron center of mass energy 
are shown in Fig.~\ref{FIG:ACTEV}. The LO and NLO results are shown in
Fig.~\ref{FIG:ACTEV}a and Fig.~\ref{FIG:ACTEV}b, respectively. Results
are displayed for the SM and two sets of anomalous couplings,
($h_{30}^Z=1.0$, $h_{40}^Z=h_{10}^Z=h_{20}^Z=0$, SM $Z\gamma\gamma$
couplings) and ($h_{40}^Z=0.05$, $h_{30}^Z=h_{10}^Z=h_{20}^Z=0$, SM 
$Z\gamma\gamma$ couplings). For simplicity, only one coupling at a time
is allowed to differ from its SM value. In order to clearly display the
effect of the anomalous couplings, we have chosen rather large values
for $h_{30}^Z$ and $h_{40}^Z$, here as well as for the LHC (see below).
However, $S$-matrix unitarity is
respected for the chosen values of the anomalous couplings, 
the power of the form factor, and
the form factor scale. The ${\cal O}(\alpha_s)$
corrections in the presence of anomalous couplings at the Tevatron energy 
are approximately 20~--~40\%, as in the SM.

For equal coupling strengths,
the numerical results obtained for the $Z\gamma\gamma$ couplings
$h_{30}^\gamma$ and $h_{40}^\gamma$ are about 20\% below those obtained
for $h_{30}^Z$ and $h_{40}^Z$ in the region where anomalous coupling
effects dominate over the SM cross section. Results for the
$CP$-violating couplings $h_{1,2}^V$ ($V=Z,\,\gamma$) are identical to
those obtained for the same values of $h_{3,4}^V$. Since terms linear in
the anomalous couplings vanish in the differential cross sections, results
are insensitive to the sign of the anomalous couplings if only one
coupling at a time is allowed to differ from its SM value.

\begin{figure}
\phantom{x}
\vskip 15cm
\includegraphics{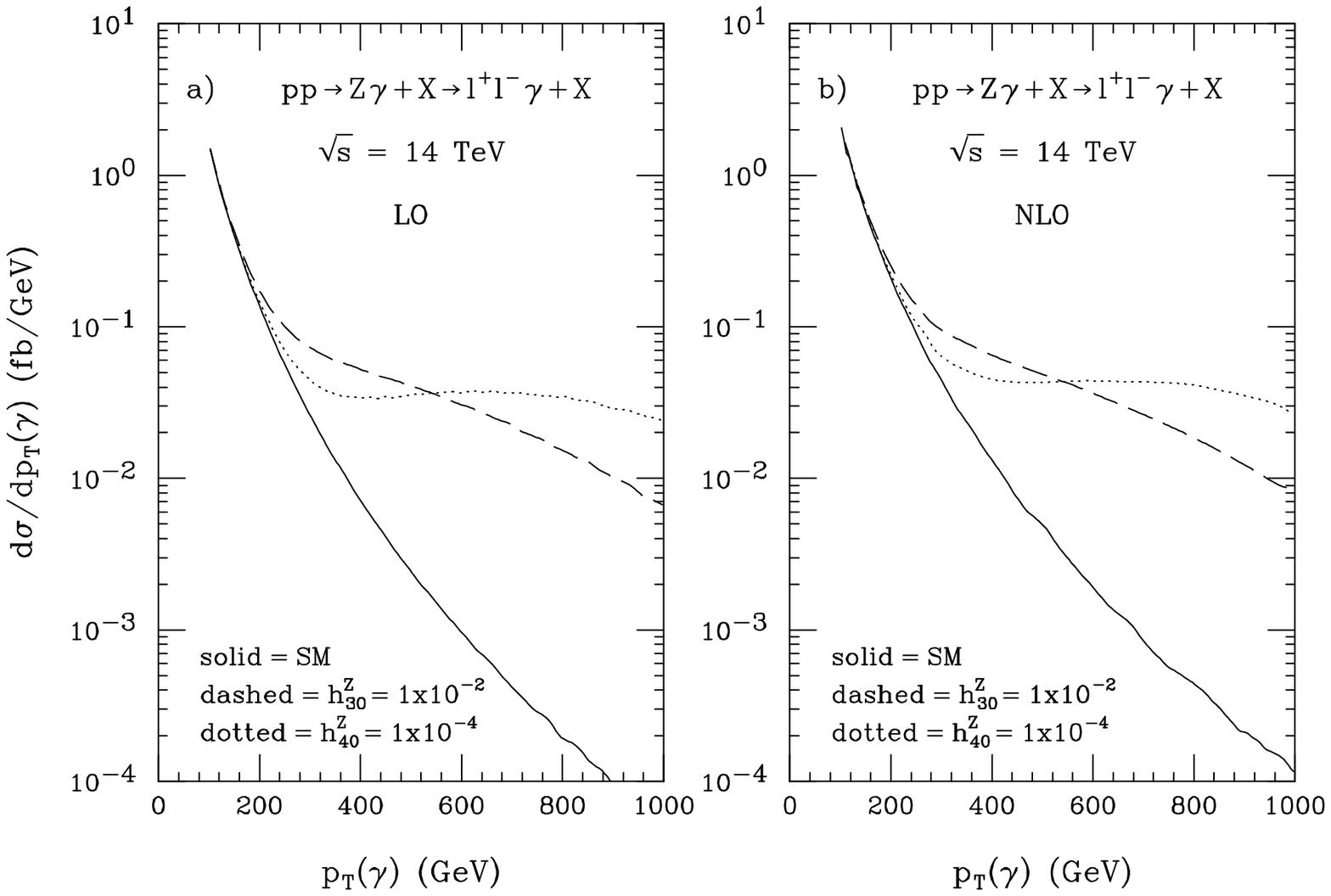}
\caption{The differential cross section for the photon transverse
momentum in the reaction $pp \to Z\gamma +X \to \ell^+ \ell^- 
\gamma + X$ at $\protect{\sqrt{s}=14}$~TeV,
(a) in the Born approximation and (b) including NLO QCD corrections. 
The curves are for the SM (solid), $h_{30}^Z=0.01$ (dashed), and 
$h_{40}^Z=1\times 10^{-4}$ (dotted). The cuts imposed are summarized in 
Sec.~IIIB.}
\label{FIG:ACLHC}
\end{figure}

The $p_T(\gamma)$ distribution for $Z(\to\ell^+\ell^-)\gamma$ production
at the LHC is shown in Fig.~\ref{FIG:ACLHC}. At leading order, the
sensitivity of the photon transverse momentum distribution to anomalous
$ZZ\gamma$ couplings is significantly  more pronounced than at the Tevatron.
In the presence of anomalous couplings, the higher order QCD corrections
are considerably smaller than in the SM. For large values of
$p_T(\gamma)$, when anomalous
couplings dominate, the ${\cal O}(\alpha_s)$ corrections are typically
between 20\% and 40\%. In the same region, QCD corrections enhance the
SM cross section by about a factor~2.2. At next-to-leading order, the
sensitivity of the photon transverse momentum spectrum to anomalous
couplings thus is somewhat reduced at the LHC.
The logarithmic factor causing the cross section enhancement at high values
of $p_T(\gamma)$ in the SM originates from the collinear region. 
The Feynman diagrams contributing in this region do not involve
the $Z\gamma V$ vertices. The logarithmic enhancement factor therefore 
does not affect the anomalous contributions to the matrix elements. 
Because $h_4^Z$ receives contributions only from operators with
dimension $\geq 8$, terms in the helicity amplitudes proportional to it
grow like $(\sqrt{s}/m_Z)^5$. Deviations originating from $h_4^Z$,
therefore, start at higher invariant masses and rise much faster than
contributions from couplings such as $h_3^Z$ which correspond to
dimension~6 operators.

\begin{figure}
\phantom{x}
\vskip 15cm
\includegraphics{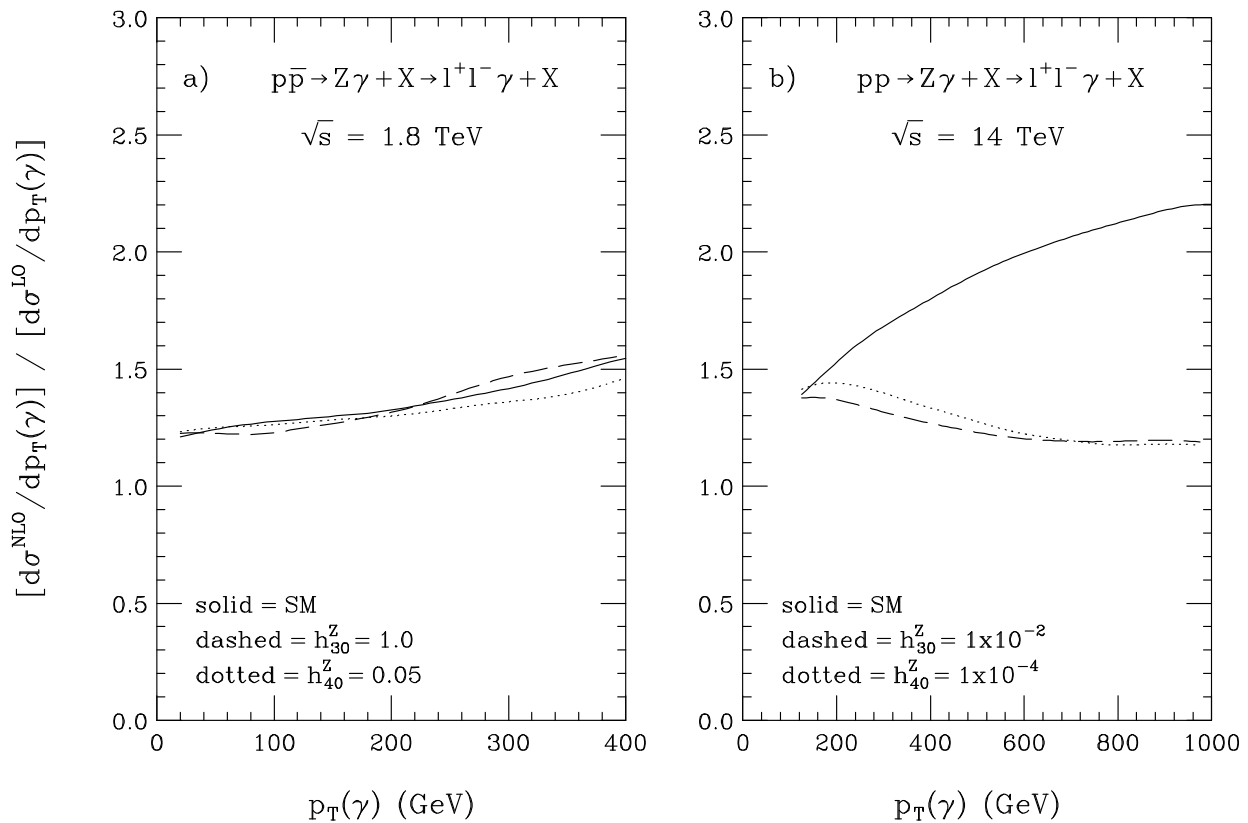}
\caption{Ratio of the NLO and LO differential cross sections of the
transverse momentum of the photon as a function of $p_T(\gamma)$ for (a)
$p\bar p\to Z\gamma +X\to\ell^+\ell^-\gamma +X$ at $\protect{\sqrt{s}=1.
8}$~TeV, and (b) $pp\to Z\gamma +X\to\ell^+\ell^-\gamma +X$ at 
$\protect{\sqrt{s}=14}$~TeV. The solid curves show the SM result. The
dashed and dotted lines display the cross section ratio for non-zero
values of $h_{30}^Z$ and $h_{40}^Z$, respectively. The cuts imposed 
are summarized in Sec.~IIIB.}
\label{FIG:RATIO}
\end{figure}

The effect of the QCD corrections is shown in more detail in 
Fig.~\ref{FIG:RATIO}, where we display the ratio of the NLO and LO
differential cross sections for the transverse momentum of the photon.
At the Tevatron, the NLO to LO cross section ratio slowly rises from 1.2
at $p_T(\gamma)=10$~GeV to about 1.5 at $p_T(\gamma)=400$~GeV. Since we
used a rather small form factor scale of $\Lambda_{FF}=500$~GeV for the
Tevatron, the effect of the anomalous couplings is suppressed at high
transverse momenta. As a result, the NLO to LO cross section ratio for
non-vanishing anomalous couplings is very similar to that obtained in
the SM. At the LHC, we use $\Lambda_{FF}=3$~TeV and the cross section
ratio gradually decreases with $p_T(\gamma)$ from $\approx 1.35$ to 1.2
if anomalous couplings are present. In contrast, the SM NLO to LO cross
section ratio increases from $\approx 1.35$ at $p_T(\gamma)=100$~GeV to
2.2 at $p_T(\gamma)=1$~TeV.

\begin{figure}
\phantom{x}
\vskip 15cm
\includegraphics{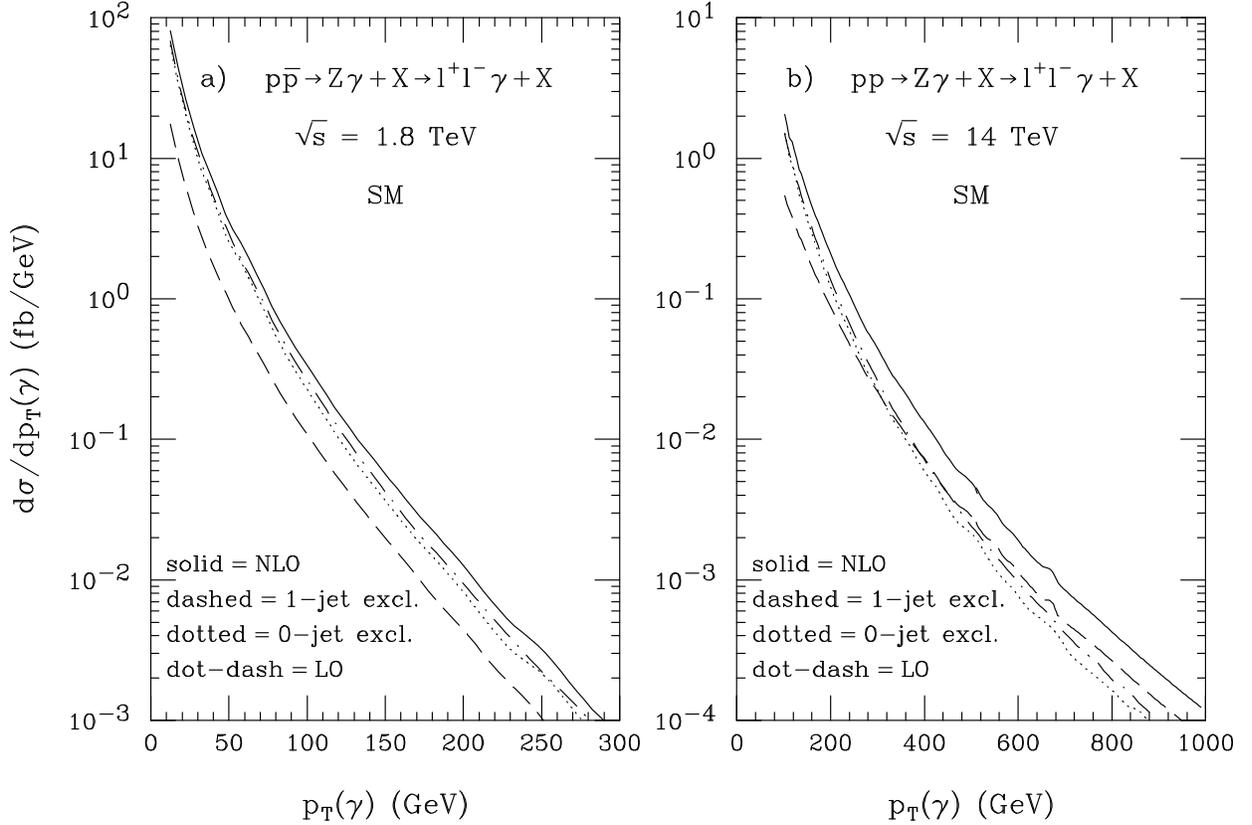}
\caption{The $p_T(\gamma)$ differential cross section for (a)
$p\bar p\to Z\gamma +X\to\ell^+\ell^-\gamma +X$ at $\protect{\sqrt{s}=1.
8}$~TeV, and (b) $pp\to Z\gamma +X\to\ell^+\ell^-\gamma +X$ at 
$\protect{\sqrt{s}=14}$~TeV in the SM. The inclusive NLO differential
cross section (solid line) is decomposed into the ${\cal O}(\alpha_s)$
0-jet (dotted line) and LO 1-jet (dashed line) exclusive
differential cross sections. For comparison, the Born cross section
(dot dashed line) is also shown. The cuts imposed are summarized in 
Sec.~IIIB. For the jet definitions, we have used 
Eqs.~(\protect{\ref{EQ:TEVJET}}) and~(\protect{\ref{EQ:LHCJET}}).}
\label{FIG:ACZEROJ}
\end{figure}

From the picture outlined above, one expects that, at next-to-leading
order, a large fraction of the $Z\gamma$ events with large photon 
transverse momentum will contain a high $p_T$ jet at the LHC. At the
Tevatron, on the other hand, the jet activity of $Z\gamma$ events at
high $p_T(\gamma)$ should be substantially reduced. This fact is
illustrated in Fig.~\ref{FIG:ACZEROJ} which shows the decomposition of
the inclusive SM NLO $p_T(\gamma)$ differential cross section into NLO
0-jet and LO 1-jet exclusive cross sections~\cite{NLOZGAMMADECAY}. For 
comparison, the photon
transverse momentum distribution in the Born approximation is also shown
in the figure. Here, a jet is defined as a quark or gluon with 
\begin{eqnarray}
p_T^{}(j)>10~{\rm GeV}\hskip 1.cm {\rm and} \hskip 1.cm |\eta(j)|<2.5
\label{EQ:TEVJET}
\end{eqnarray}
at the Tevatron, and 
\begin{eqnarray}
p_T^{}(j)>50~{\rm GeV}\hskip 1.cm {\rm and} \hskip 1.cm |\eta(j)|<3
\label{EQ:LHCJET}
\end{eqnarray}
at the LHC. The sum of the NLO 0-jet and the LO 1-jet exclusive cross 
section is equal to the inclusive NLO cross section. 
The NLO exclusive $Z\gamma + 0$~jet and the LO exclusive
$Z\gamma + 1$~jet cross sections depend explicitly on the jet
definition, however, the inclusive NLO cross section is independent of the
jet definition. 

Present LHC studies~\cite{CMS,ATLAS,LHCJET} and projections to Tevatron
energies suggest that jets fulfilling the
criteria of Eqs.~(\ref{EQ:TEVJET}) and~(\ref{EQ:LHCJET}) can be 
identified in $Z\gamma+X$ events at the Tevatron~\cite{GPJ} and 
LHC~\cite{LHC} for luminosities up to $10^{33}$~cm$^{-2}$~s$^{-1}$ and 
$10^{34}$~cm$^{-2}$~s$^{-1}$, respectively. It should be noted,
however, that for theoretical reasons, the jet transverse momentum 
threshold can not be made 
arbitrarily small in our calculation. For transverse momenta below
5~GeV (20~GeV) at the Tevatron (LHC), soft gluon resummation effects
are expected to significantly change the shape of the jet $p_T^{}$ 
distribution~\cite{RESUM}. For the jet definitions discussed above,
these effects are expected to be unimportant and therefore are ignored
in our calculation. 

\begin{figure}
\phantom{x}
\vskip 15cm
\includegraphics{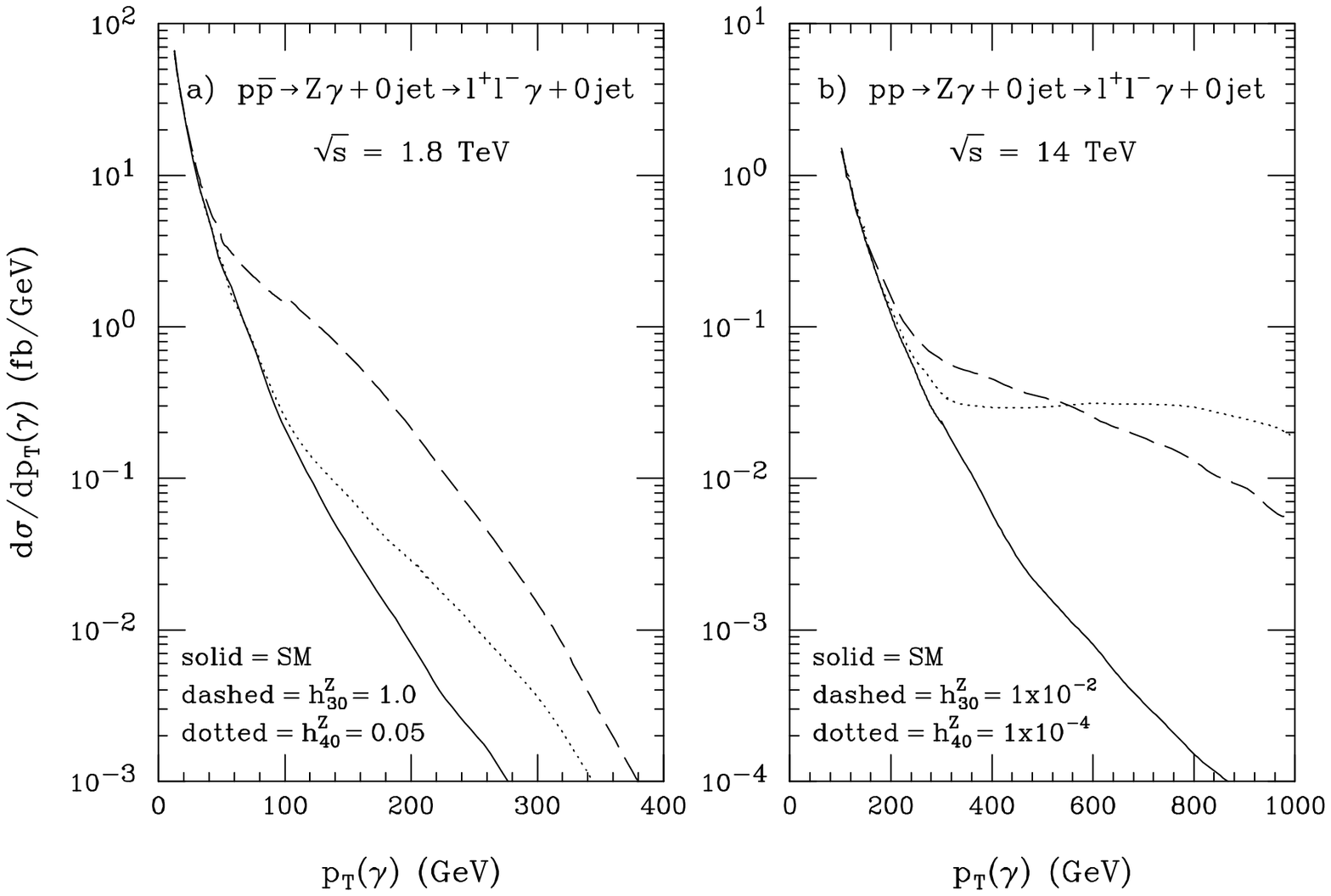}
\caption{The $p_T(\gamma)$ differential cross section for (a)
$p\bar p\to Z\gamma +0~{\rm jet}\to\ell^+\ell^-\gamma +0~{\rm jet}$ at 
$\protect{\sqrt{s}=1.8}$~TeV, and (b) $pp\to Z\gamma +0~{\rm 
jet}\to\ell^+\ell^-\gamma +0~{\rm jet}$ at $\protect{\sqrt{s}=14}$~TeV. 
The curves in part (a) are for the SM (solid), $h_{30}^Z=1.0$ (dashed), and 
$h_{40}^Z=0.05$ (dotted). In part (b), the dashed and dotted curves are
for $h_{30}^Z=10^{-2}$ and $h_{40}^Z=10^{-4}$, respectively. 
The cuts imposed are 
summarized in Sec.~IIIB. For the jet definitions, we have used 
Eqs.~(\protect{\ref{EQ:TEVJET}}) and~(\protect{\ref{EQ:LHCJET}}).}
\label{FIG:NSMZEROJ}
\end{figure}

Figure~\ref{FIG:ACZEROJ} shows that, at the Tevatron, the 1-jet cross
section is always considerably smaller than the NLO 0-jet rate. At the 
LHC, on the
other hand, the 1-jet cross section is larger than the LO cross
section for $p_T(\gamma)> 400$~GeV. The effect of the QCD corrections
can be reduced by vetoing hard jets in the central rapidity region, {\it
i.e.}, by imposing a ``zero jet'' requirement and considering
$Z\gamma+0$~jet production only. The NLO 0-jet and Born differential
cross sections deviate by 30\% at most in the $p_T$ region shown. 
The photon transverse momentum distribution for NLO $Z\gamma+0$~jet
production is shown in Fig.~\ref{FIG:NSMZEROJ}. The 0-jet requirement is
seen to restore the sensitivity to anomalous couplings lost
in the inclusive NLO cross section at the LHC. It has little effect at
the Tevatron. 

As mentioned in Sec.~IIIA, all our results are obtained for a scale of
$Q^2=M^2_{Z\gamma}$. The Born cross section for $Z\gamma$ production depends
significantly on the choice of $Q$, which enters through the 
scale-dependence of the parton distribution functions. At the NLO level, 
the  $Q$-dependence enters not only via the 
parton distribution functions, but also through the running coupling 
$\alpha_s(Q^2)$ and the explicit factorization scale-dependence in the 
order $\alpha_s(Q^2)$ correction terms. Similar to the situation
encountered in $W\gamma$, $WZ$, and $W^+W^-$ production in hadronic
collisions~\cite{BHO,BHO1,BHO2}, we find that the NLO $Z\gamma+0$~jet 
exclusive cross section is almost independent of the scale $Q$. Here, 
the scale-dependence of the
parton distribution functions is compensated by that of $\alpha_s(Q^2)$ 
and the explicit factorization scale dependence in the correction terms.
The $Q$-dependence of the inclusive NLO cross section is 
larger than that of the NLO 0-jet cross section; it is dominated by the 
1-jet exclusive component which is calculated only to lowest order and
thus exhibits a considerable scale-dependence. 

\subsection{${\cal O} (\alpha_s)$ Corrections and
Anomalous $ZZ\gamma$ and $Z\gamma\gamma$ Couplings in
$Z(\to\bar\nu\nu)\gamma$ Production}

If the $Z$ boson produced in $q\bar q\to Z\gamma$ decays into neutrinos, the
signal consists of a high $p_T$ photon accompanied by a large amount of
missing transverse momentum, $p\llap/_T$. Due to
the larger $Z\to\bar\nu\nu$ branching ratio, the $\gamma p\llap/_T+X$ 
differential cross section is about a factor~3 larger than that for 
$q\bar q\to e^+e^-\gamma+X$ and $q\bar q\to\mu^+\mu^-\gamma+X$ combined.
This results in limits on the anomalous $ZZ\gamma$ and $Z\gamma\gamma$
couplings which are presently about a factor~2 better than those 
obtained from the $Z(\to\ell^+\ell^-)\gamma$ analysis~\cite{d0nndata}. 

\begin{figure}
\phantom{x}
\vskip 15cm
\includegraphics{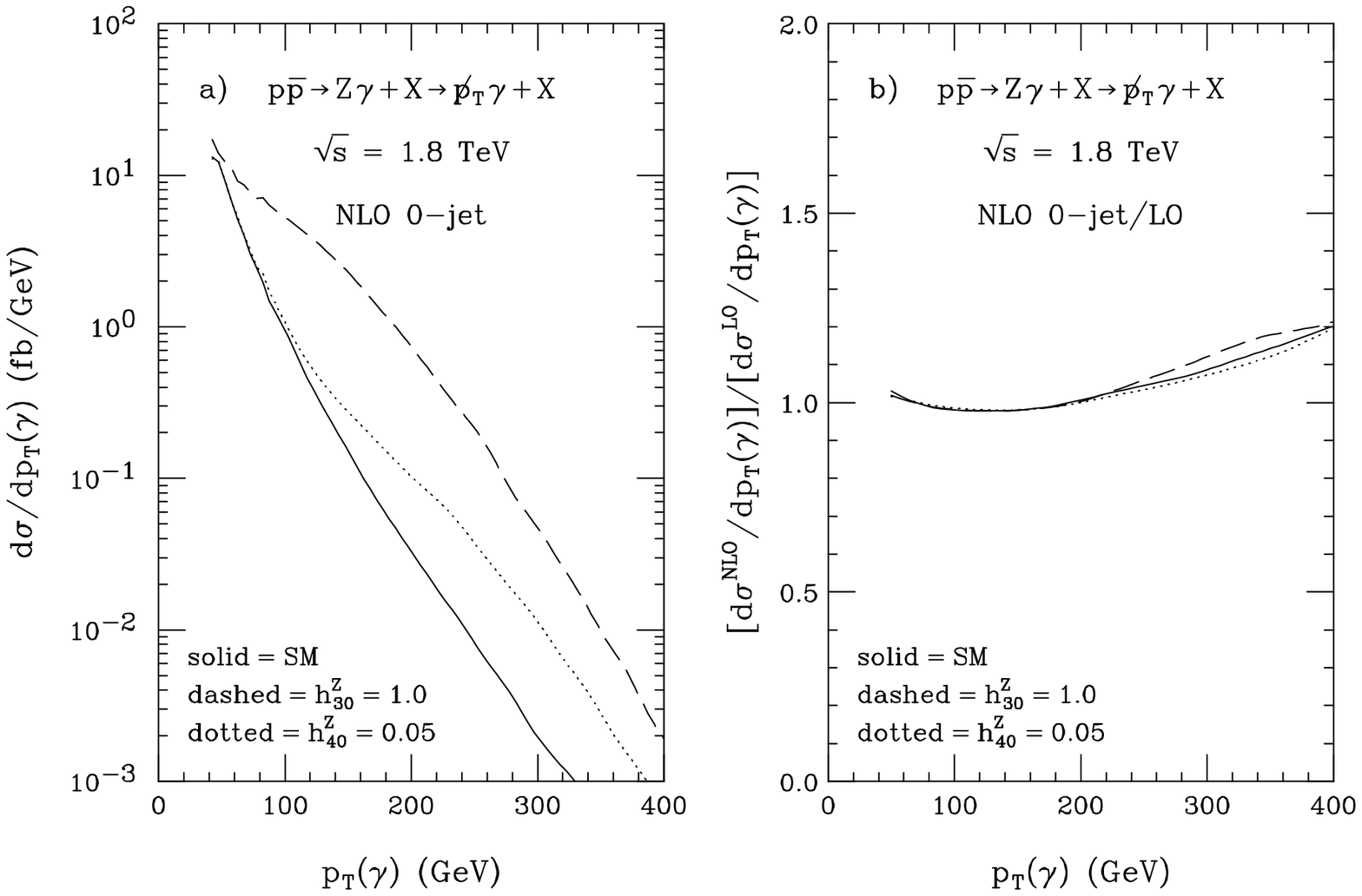}
\caption{(a) The $p_T(\gamma)$ differential cross section, and (b) the 
ratio of the NLO 0-jet to LO differential cross sections of the photon 
transverse momentum as a function of $p_T(\gamma)$ for 
$p\bar p\to Z\gamma +0~{\rm jet}\to p\llap/_T\gamma +0~{\rm jet}$ at 
$\protect{\sqrt{s}=1.8}$~TeV.
The curves are for the SM (solid), $h_{30}^Z=1.0$ (dashed), and 
$h_{40}^Z=0.05$ (dotted). The cuts imposed are summarized in Sec.~IIIB. }
\label{FIG:TEVNU}
\end{figure}

The NLO photon transverse momentum distribution for $Z(\to\bar\nu\nu)\gamma$
production at the Tevatron is shown in Fig.~\ref{FIG:TEVNU}a. Here we
have imposed the cuts of the D\O\ $Z(\to\bar\nu\nu)\gamma$ analysis (see
Sec. IIIB and Ref.~\cite{d0nndata}). Since jets with a transverse energy
larger than 15~GeV are excluded in the experimental analysis, only the
0-jet differential cross section is shown. The effect of non-standard 
$Z\gamma V$ couplings is very similar to that observed in
$Z(\to\ell^+\ell^-)\gamma$ production. The impact of the QCD
corrections on the differential cross section is shown in 
Fig.~\ref{FIG:TEVNU}b, where we display the ratio of the NLO and LO 
differential cross sections for the transverse momentum of the photon.
Due to the jet veto cut imposed, the QCD corrections are small over a
wide range of photon transverse momenta. For $p_T(\gamma)<200$~GeV, the cross
section ratio is almost constant. It rises slowly for larger values of
$p_T(\gamma)$. The NLO to LO cross section ratio in
the SM and in the presence of anomalous couplings are very similar. 

Figure~\ref{FIG:TEVNU}b demonstrates that, for data samples containing
only a few $Z(\to\bar\nu\nu)\gamma$ events with $p_T(\gamma)>200$~GeV and
for the jet veto cut used, the LO calculation is an adequate
approximation. However, increasing or decreasing the jet transverse
momentum threshold will change the size of the QCD corrections, and thus
increase the deviation of the NLO calculation from the LO result. The
calculation of Ref.~\cite{BAURBERGER}, where the effect of the NLO QCD
corrections is approximated by a constant $k$-factor of
$k=1+8\alpha_s/9\pi\approx 1.34$, overestimates the cross section by
about 30\% over a wide range of photon transverse momenta. This
calculation has been used in the D\O\ analysis to compare data with the SM
prediction, and to extract limits for the $ZZ\gamma$ and $Z\gamma\gamma$
couplings. Since it overestimates the cross section, the limits obtained
are slightly better than those extracted using the full NLO calculation
(see Sec.~IIIF).

\begin{figure}
\phantom{x}
\vskip 15cm
\includegraphics{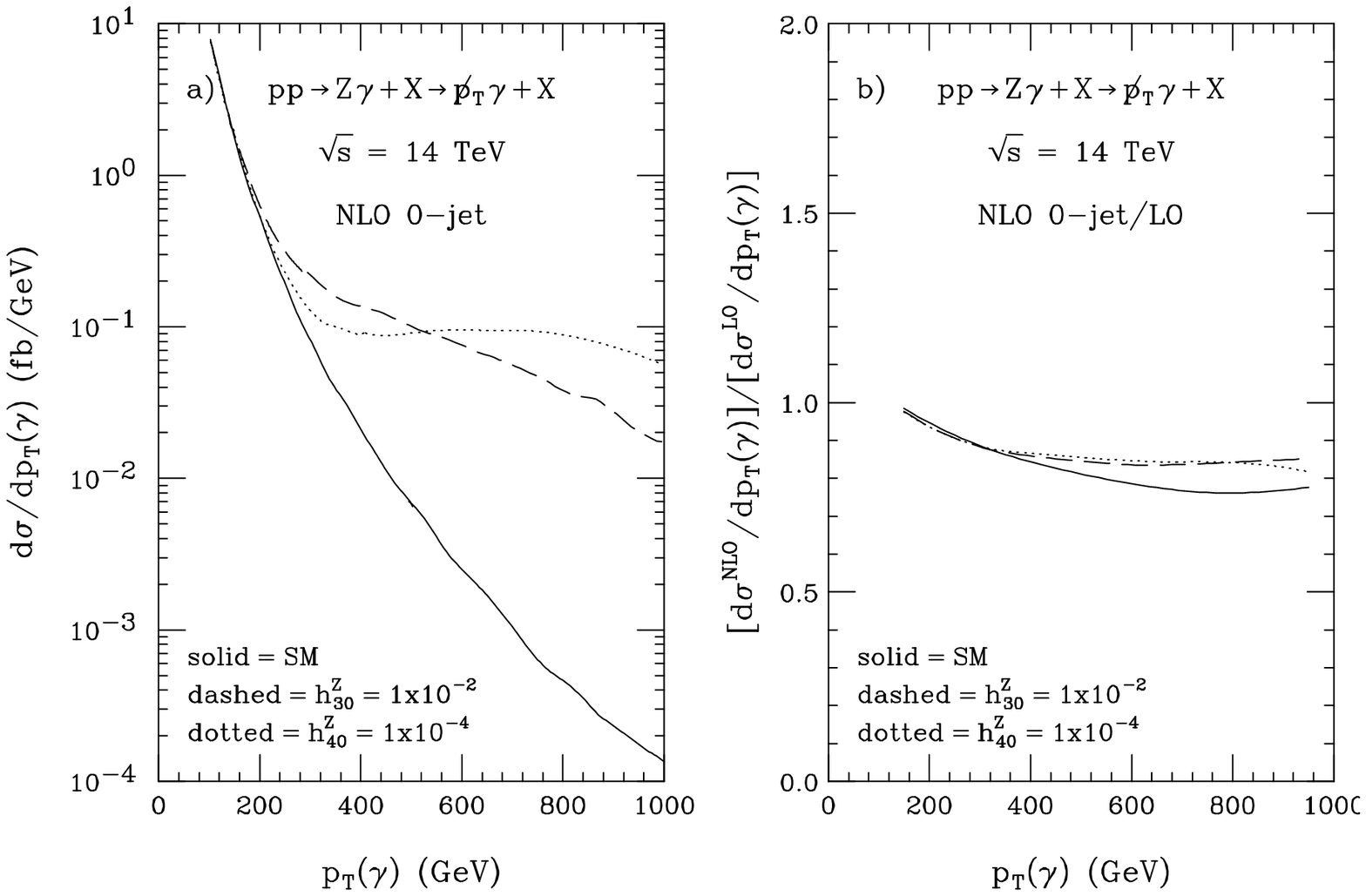}
\caption{(a) The $p_T(\gamma)$ differential cross section, and (b) the 
ratio of the NLO 0-jet to LO differential cross sections of the photon 
transverse momentum as a function of $p_T(\gamma)$ for 
$pp\to Z\gamma +0~{\rm jet}\to p\llap/_T\gamma +0~{\rm jet}$ at 
$\protect{\sqrt{s}=14}$~TeV.
The curves are for the SM (solid), $h_{30}^Z=10^{-2}$ (dashed), and 
$h_{40}^Z=10^{-4}$ (dotted). The cuts imposed are summarized in Sec.~IIIB. }
\label{FIG:LHCNU}
\end{figure}

In Fig.~\ref{FIG:LHCNU}a we show the NLO photon transverse momentum
distribution for $Z(\to\bar\nu\nu)\gamma$ production at the LHC,
imposing a $p_T(j)<50$~GeV jet veto cut. Figure~\ref{FIG:LHCNU}b
displays the ratio of the NLO and LO differential cross sections for
the transverse momentum of the photon. For the $p_T(j)$ threshold
chosen, NLO QCD corrections reduce the cross section by up to 20\%. In
contrast to the situation encountered at the Tevatron, the
cross section ratio slowly falls with $p_T(\gamma)$.

\subsection{Sensitivity Limits}

We now study the impact that ${\cal O}(\alpha_s)$ QCD corrections to
$Z\gamma$ production have on the sensitivity limits for $h_{i0}^V$
at the Tevatron and LHC.  For the Tevatron we
consider integrated luminosities of 1~fb$^{-1}$, as envisioned for the 
Main Injector era, and 10~fb$^{-1}$ (TeV33) which could be achieved
through additional upgrades of the Tevatron accelerator
complex~\cite{GPJ}. In the case of the LHC we use $\int\!{\cal
L}dt=10$~fb$^{-1}$ and 100~fb$^{-1}$~\cite{LHC}. To extract 
limits in the $Z(\to\ell^+\ell^-)\gamma$ case, we sum over electron and
muon final states. Interference effects between different $Z\gamma V$
$(V = Z, \gamma)$ couplings are fully incorporated in our analysis. 

To derive 95\%~CL limits we use the $p_T(\gamma)$
distribution and perform a $\chi^2$ test~\cite{BAURBERGER}, assuming that no 
deviations from the SM predictions are observed in the experiments
considered. We include the cuts summarized in Sec.~IIIB. For 
$Z(\to\ell^+\ell^-)\gamma$ production, we use the jet definitions of
Eqs.~(\ref{EQ:TEVJET}) and~(\ref{EQ:LHCJET}). Unless explicitly stated 
otherwise, a form factor as given in Eq.~(\ref{EQ:DIPOLES}) is used with
$n=3$ for $h_{1,3}^V$, and $n=4$ for $h_{2,4}^V$. Furthermore, the form 
factor scale $\Lambda_{FF}$ is taken to be 0.5~TeV (3.0~TeV) for 
Tevatron (LHC) simulations. The $p_T(\gamma)$
distribution is split into a certain number of bins. The number of bins
and the bin width depend on
the center of mass energy and the integrated luminosity. In each bin the
Poisson statistics are approximated by a Gaussian distribution. In order
to achieve a sizable counting rate in each bin, all events above a
certain threshold are collected in a single bin. 
This procedure guarantees that a high statistical significance cannot
arise from a single event at large transverse momentum, where the
SM predicts much less than one event. In order to derive
realistic limits we allow for a normalization uncertainty of 50\% in the
SM cross section. For the cuts we impose, background contributions other
than SM $Z\gamma$ production are small~\cite{BAURBERGER} and are ignored
in our derivation of sensitivity limits.

The calculation of sensitivity bounds is facilitated by the observation that
the $CP$ conserving couplings $h_{3,4}^V$ and the $CP$ violating
couplings $h_{1,2}^V$ do not interfere. Furthermore, cross sections and
sensitivities are
nearly identical for equal values of $h_{10,20}^V$ and $h_{30,40}^V$. In the
following we shall therefore concentrate on $h_{3,4}^V$. In each bin,
$i$, the cross section is a bilinear function of the anomalous
couplings:
\begin{equation}
\sigma^i=\sigma^i(SM)+\sum_{V=\gamma,\,Z} \left (a^{iV}_3\,h_3^V + 
a^{iV}_4\,h_4^V\right ) + \sum_{V,V'=\gamma,\,Z} \sum_{j,k=3,4}
b^{iVV'}_{jk}\,h_j^V\,h_k^{V'}.
\label{EQ:SENS}
\end{equation}
Here, $\sigma^i(SM)$ is the SM cross section, and $a^{iV}_{3,4}$ and
$b^{iVV'}_{jk}$ are constants. Since the interference terms between the
SM and the anomalous contributions to the helicity amplitudes vanish for
$h_{3,4}^V$ (see Sec.~IIB), 
\begin{equation}
a^{iV}_3=a^{iV}_4=0.
\end{equation}
These constraints are taken into account in our calculation of sensitivity
bounds. 

For $h^V_{1,2}$, an expression for the cross section similar to that of
Eq.~(\ref{EQ:SENS}) can be derived. In this case, however, the
coefficients of the terms linear in $h^V_{1,2}$ only vanish if the phase
space integration over the scattering angle of the photon, $\theta$, is
symmetric in $\cos\theta$ (see Sec.~IIB).

\begin{figure}
\phantom{x}
\vskip 15cm
\includegraphics{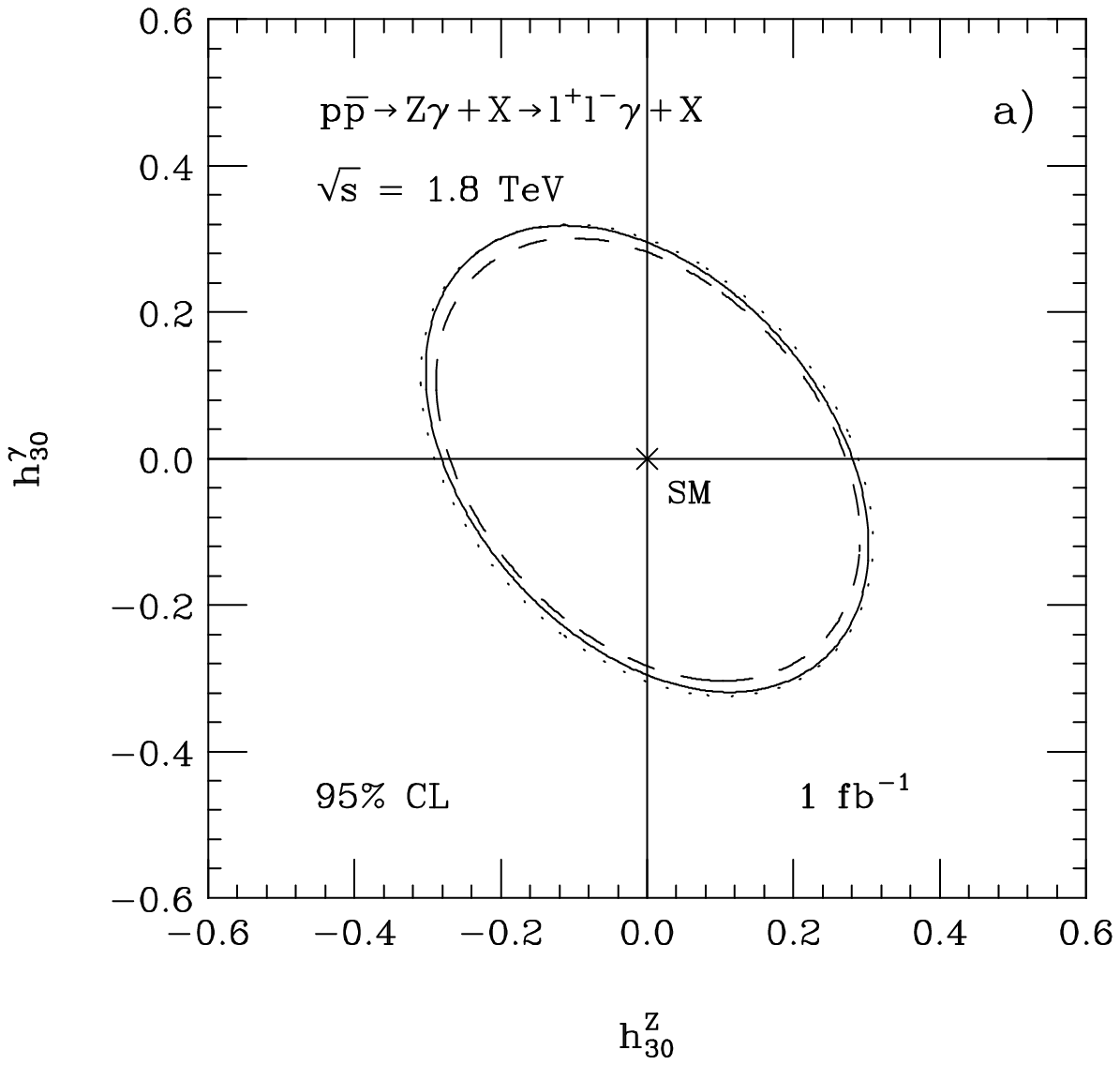}
\includegraphics{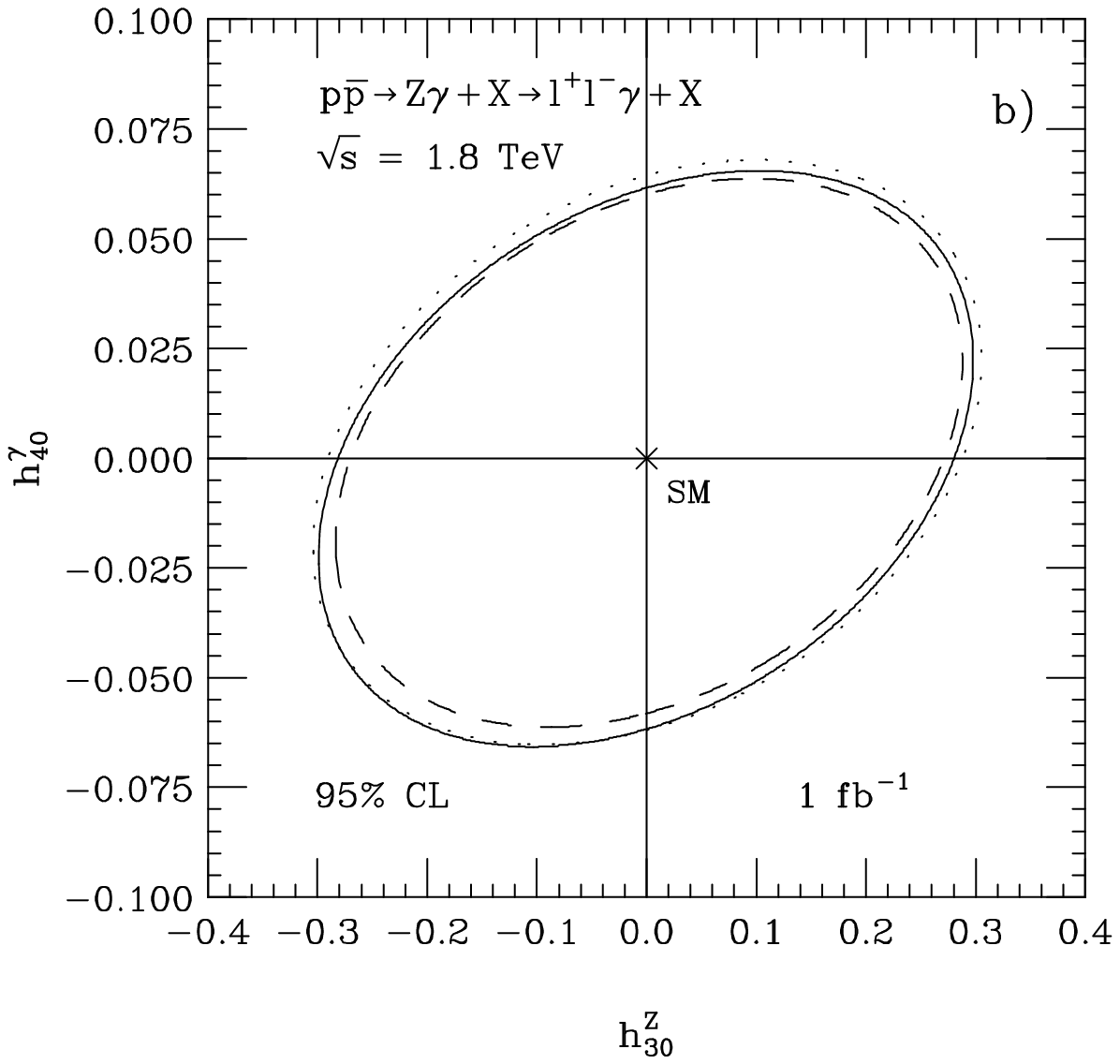}
\includegraphics{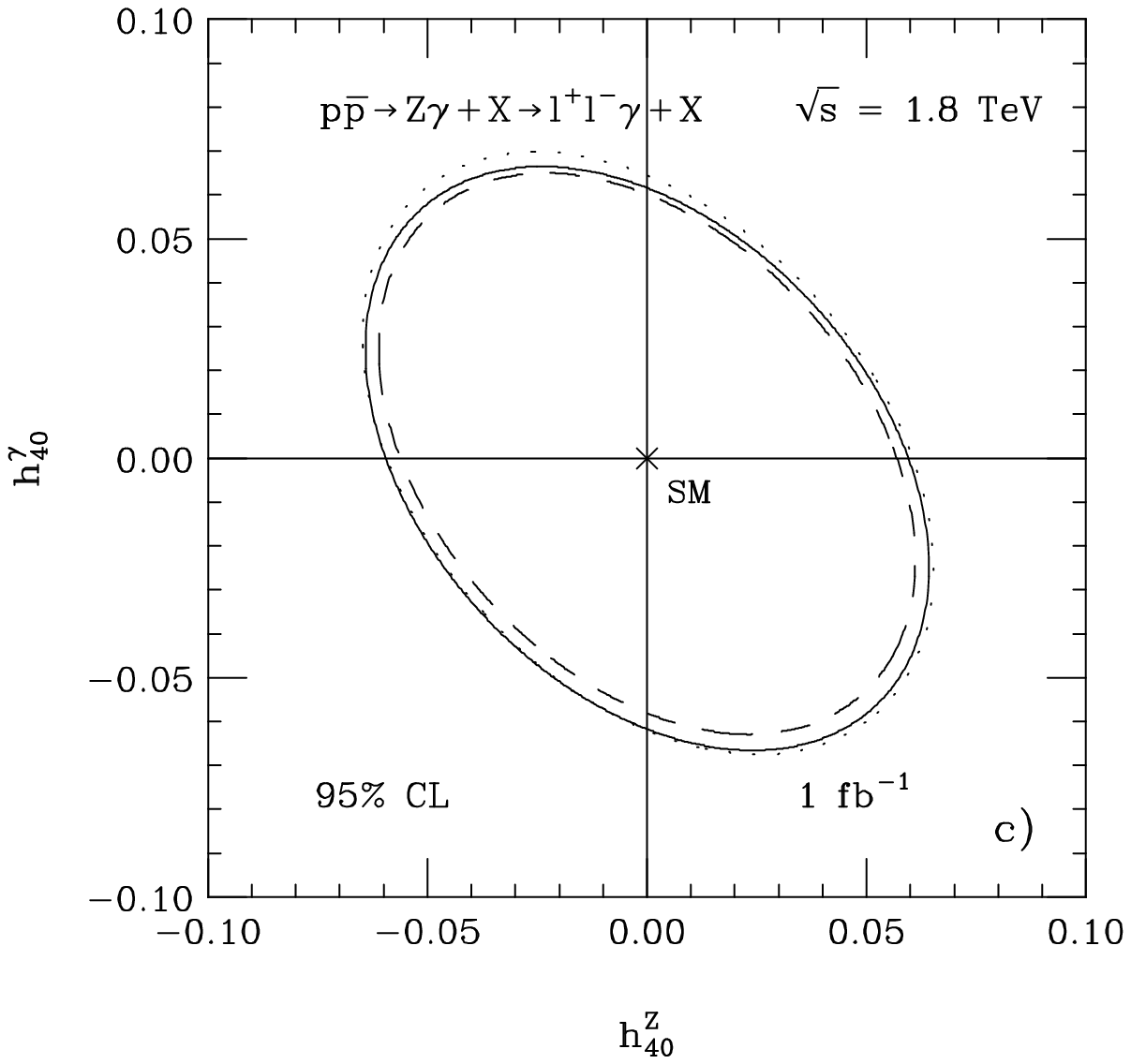}
\caption{Limit contours at the \protect{95\%~CL} for 
\protect{$p\bar p\rightarrow
Z\gamma+X\rightarrow\ell^+\ell^-\gamma+X$} ($\ell=e,\,\mu$), 
derived from the $p_T(\gamma)$ distribution at the Tevatron for
$\int\!{\cal L}dt=1$~fb$^{-1}$. Contours are shown in three planes:
a) the \protect{$h_{30}^Z$} \protect{--}~\protect{$h_{30}^\gamma$} 
plane, b) the \protect{$h_{30}^Z$} \protect{--}~\protect{$h_{40}^\gamma$}
plane, and c) the \protect{$h_{40}^Z$}
\protect{--}~\protect{$h_{40}^\gamma$} 
plane. The solid lines give the results for LO $Z\gamma$ production.
The dashed curves give the inclusive NLO results, and the dotted lines
show the bounds obtained from the exclusive $Z\gamma+0$~jet channel. 
In each graph, only those couplings
which are plotted against each other are assumed to be different from 
their zero SM values. The
cuts imposed are summarized in Sec.~IIIB. For the jet definition,
we have used Eq.~(\protect{\ref{EQ:TEVJET}}). }
\label{FIG:TWELVE}
\end{figure}

\begin{figure}
\phantom{x}
\vskip 15cm
\includegraphics{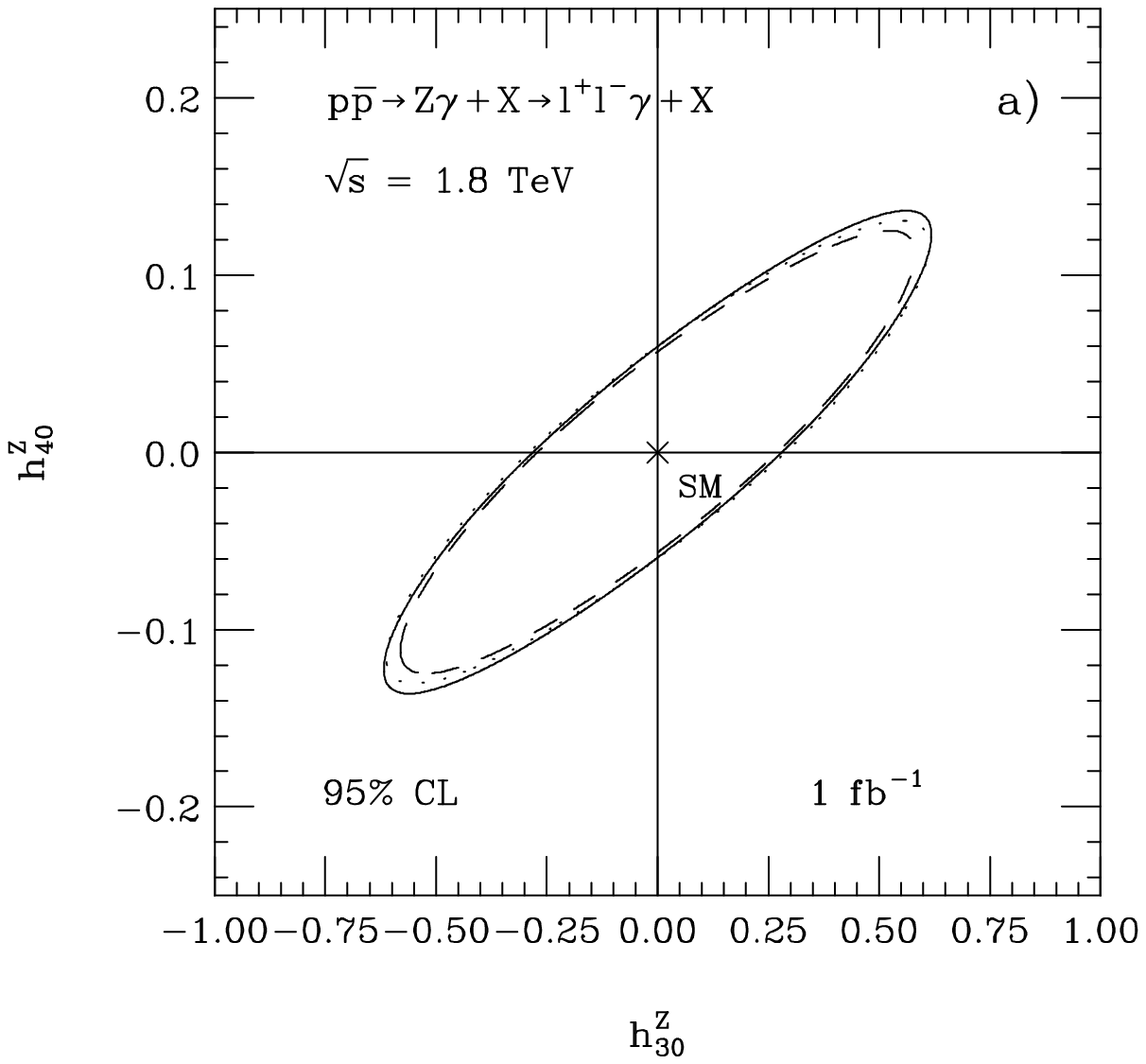}
\includegraphics{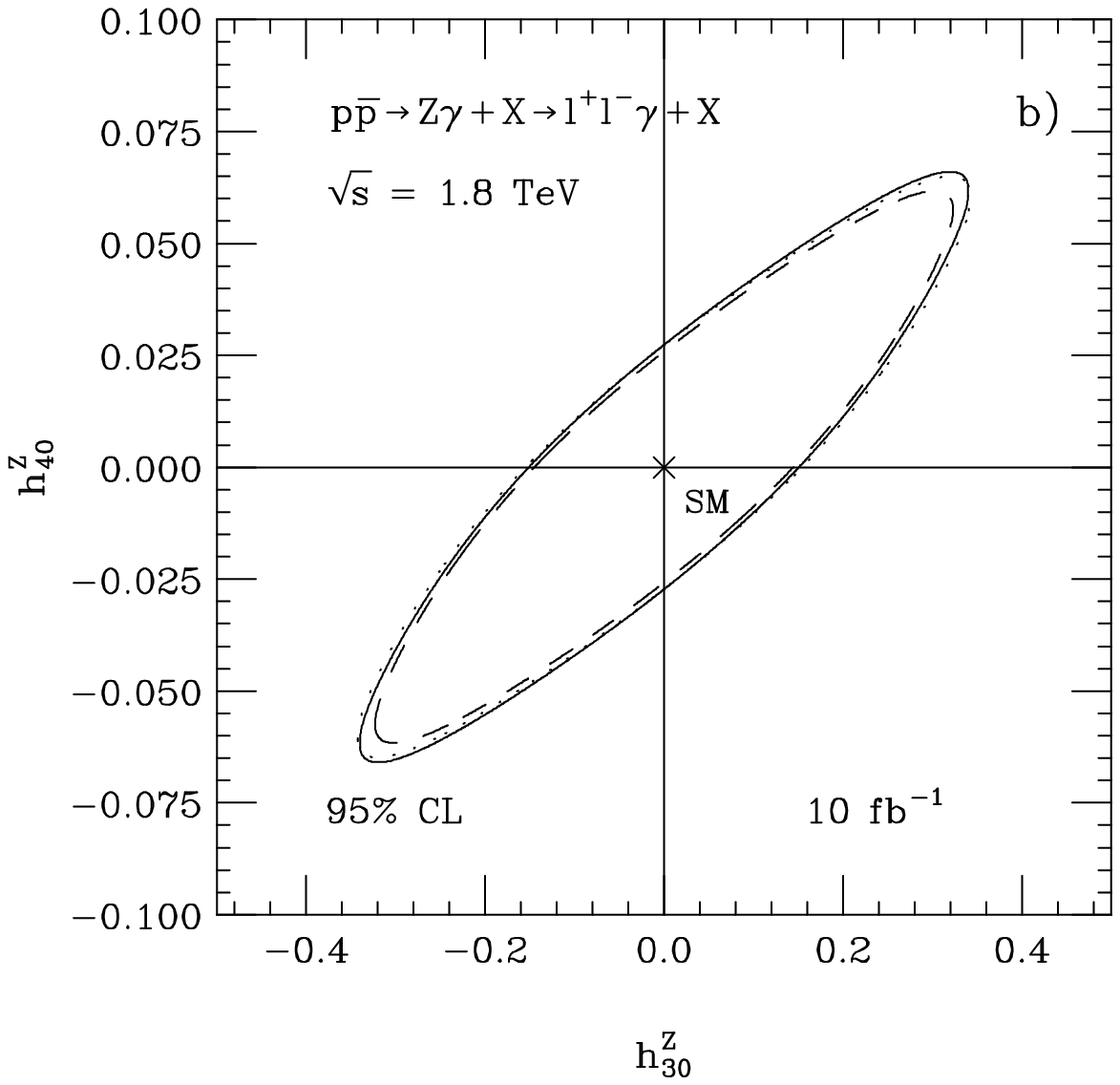}
\caption{Limit contours at the \protect{95\%~CL} in the
$h_{30}^Z - h_{40}^Z$ plane for \protect{$p\bar p\rightarrow
Z\gamma+X\rightarrow\ell^+\ell^-\gamma+X$} ($\ell=e,\,\mu$), 
derived from the $p_T(\gamma)$ distribution at the Tevatron for a)
$\int\!{\cal L}dt=1$~fb$^{-1}$ and b) $\int\!{\cal L}dt=10$~fb$^{-1}$. 
The solid lines give the results for LO $Z\gamma$ production.
The dashed curves give the inclusive NLO results, and the dotted lines
show the bounds obtained from the exclusive $Z\gamma+0$~jet channel. 
$h_{1,2}^Z$ and all $Z\gamma\gamma$ couplings are assumed to be zero.
The cuts imposed are summarized in Sec.~IIIB. For the jet definition,
we have used Eq.~(\protect{\ref{EQ:TEVJET}}). }
\label{FIG:THIRTEEN}
\end{figure}

\begin{figure}
\phantom{x}
\vskip 15cm
\includegraphics{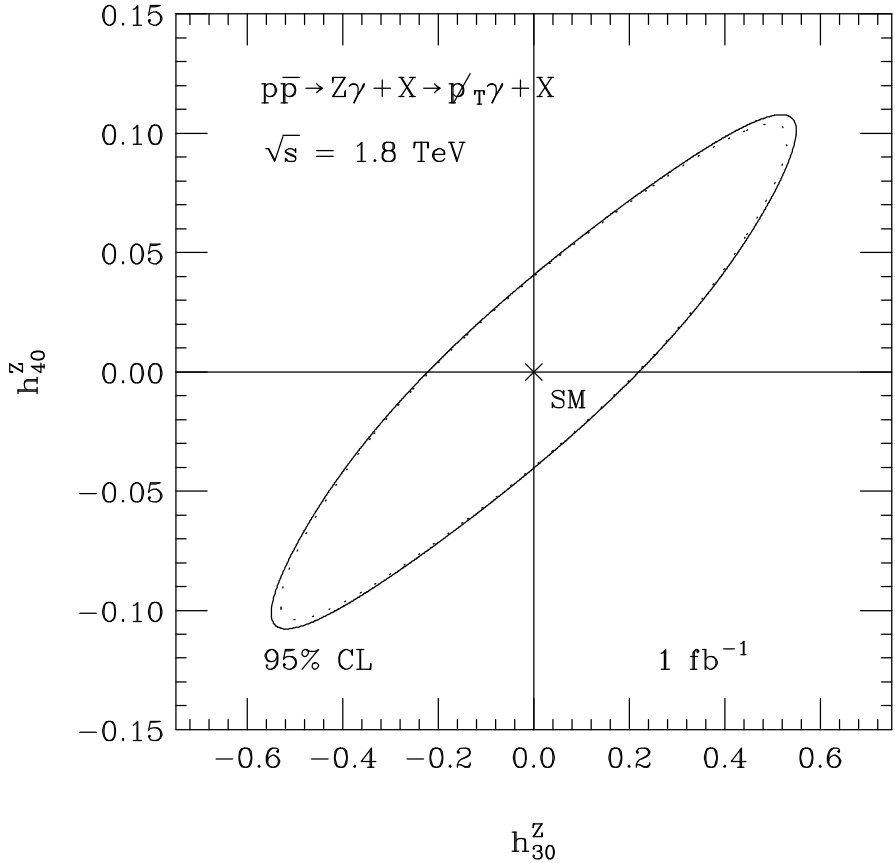}
\caption{Limit contours at the \protect{95\%~CL} in the
$h_{30}^Z - h_{40}^Z$ plane for \protect{$p\bar p\rightarrow
Z\gamma+X\rightarrow p\llap/_T\gamma+X$} derived from the $p_T(\gamma)$ 
distribution at the Tevatron for $\int\!{\cal L}dt=1$~fb$^{-1}$. 
The solid line gives the result for LO $Z\gamma$ production.
The dotted curve shows the bounds obtained from the NLO calculation.
$h_{1,2}^Z$ and all $Z\gamma\gamma$ couplings are assumed to be zero.
The cuts imposed are summarized in Sec.~IIIB. }
\label{FIG:FOURTEEN}
\end{figure}

\begin{figure}
\phantom{x}
\vskip 15cm
\includegraphics{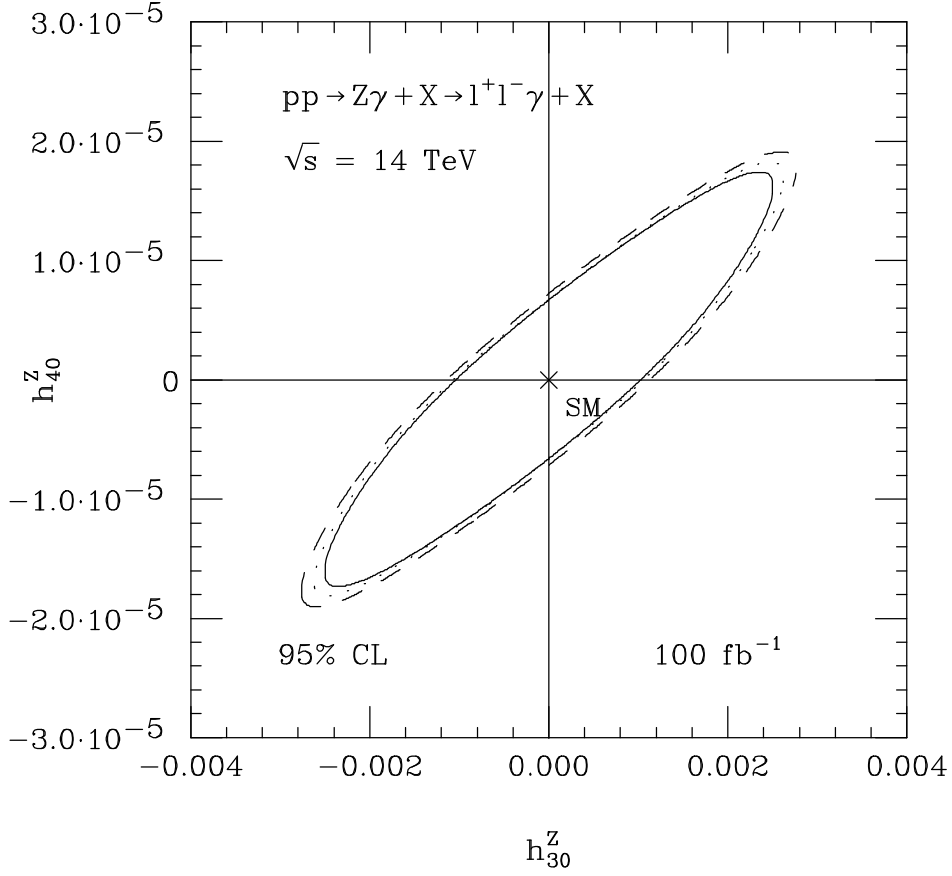}
\caption{Limit contours at the \protect{95\%~CL} in the
$h_{30}^Z - h_{40}^Z$ plane for \protect{$pp\rightarrow
Z\gamma+X\rightarrow\ell^+\ell^-\gamma+X$} ($\ell=e,\,\mu$) derived
from the $p_T(\gamma)$ 
distribution at the LHC for $\int\!{\cal L}dt=100$~fb$^{-1}$. 
The solid line gives the results for LO $Z\gamma$ production.
The dashed curve gives the inclusive NLO results, and the dotted lines
shows the bounds obtained from the exclusive $Z\gamma+0$~jet channel. 
$h_{1,2}^Z$ and all $Z\gamma\gamma$ couplings are assumed to be zero.
The cuts imposed are summarized in Sec.~IIIB. For the jet definition,
we have used Eq.~(\protect{\ref{EQ:LHCJET}}). }
\label{FIG:FIFTEEN}
\end{figure}

Our results are summarized in Figs.~\ref{FIG:TWELVE} 
--~\ref{FIG:FIFTEEN} and Tables~\ref{TAB:ONE} --~\ref{TAB:THREE}.
Figure~\ref{FIG:TWELVE} shows 95\%~CL contours in the $h_{30}^Z -
h_{30}^\gamma$, $h_{30}^Z - h_{40}^\gamma$, and the $h_{40}^Z -
h_{40}^\gamma$ plane for $Z(\to\ell^+\ell^-)\gamma$ production and an
integrated luminosity of 1~fb$^{-1}$ at the Tevatron. Results for the 
$h_{30}^\gamma - h_{40}^Z$ plane are very similar to those displayed in 
Fig.~\ref{FIG:TWELVE}b for $h_{30}^Z$ versus $h_{40}^\gamma$, and are 
therefore not shown. In each figure, only those couplings
which are plotted against each other are assumed to be different from 
their zero SM values. As noted in Ref.~\cite{BAURBERGER}, $ZZ\gamma$ and
$Z\gamma\gamma$ couplings interfere little at LO. 
Figure~\ref{FIG:TWELVE} demonstrates that the NLO QCD corrections do not
change this behaviour. This statement also applies to different
integrated luminosities and to $Z(\to\ell^+\ell^-)\gamma$ production at
the LHC. Due to the larger cross section, the sensitivity bounds
obtained from the inclusive NLO differential cross section are about 5\%
better than those derived using the LO calculation. The limits extracted
from the NLO $Z\gamma+0$~jet cross section are almost identical to those
found using the LO calculation. 

Larger correlations are encountered between $h_3^V$ and $h_4^V$ (see
Ref.~\cite{BAURBERGER}). The impact of the ${\cal O}(\alpha_s)$ QCD
corrections on the limits in the $h_3^Z - h_4^Z$ plane at the Tevatron 
is shown in 
Fig.~\ref{FIG:THIRTEEN} for integrated luminosities of 1~fb$^{-1}$ and
10~fb$^{-1}$. The inclusive NLO QCD corrections improve the
sensitivity bounds by up to 9\% (6\%) for 1~fb$^{-1}$ (10~fb$^{-1}$)
whereas the limits obtained analyzing the $Z\gamma+0$~jet channel
are very similar to those found using the LO calculation. 
Similar results are obtained for $h_3^\gamma$ and $h_4^\gamma$. The
sensitivity bounds for the $Z\gamma\gamma$ couplings are a few per cent
weaker than those found for the corresponding $ZZ\gamma$
couplings, and QCD corrections have a smaller effect than in the case of
the $ZZ\gamma$ couplings. Table~\ref{TAB:ONE} summarizes the 95\%~CL
sensitivity limits, including all correlations, at leading order and
next-to-leading order for $h_{30}^V$ and $h_{40}^V$ ($V=Z,\,\gamma$) for
the process $p\bar p\to Z\gamma+X\to\ell^+\ell^-\gamma+X$ at the
Tevatron with $\int\!{\cal L}dt=1~{\rm fb}^{-1}$.

Figure~\ref{FIG:FOURTEEN} displays how the 95\%~CL contour limits for
$p\bar p\to Z\gamma +X\to p\llap/_T\gamma+X$ at the Tevatron and an
integrated luminosity of 1~fb$^{-1}$ are affected by NLO QCD
corrections. Since a jet veto cut of $p_T(j)<15$~GeV is imposed, we
only show the LO (solid line) and NLO 0-jet
(dotted line) contour limits. As we have demonstrated in Sec.~IIIE, the
NLO QCD corrections to $Z(\to\bar\nu\nu)\gamma$ production at the
Tevatron are small over a large range of photon transverse momenta due
to the 0-jet requirement. The sensitivities achievable for $h_{30}^Z$
and $h_{40}^Z$ in $p\bar p\to Z\gamma +X\to p\llap/_T\gamma+X$ at the 
Tevatron for $\int\!{\cal L}dt=1~{\rm fb}^{-1}$ and $\int\!{\cal 
L}dt=10~{\rm fb}^{-1}$ are listed in Table~\ref{TAB:TWO}. Besides the LO
and NLO 0-jet bounds, we also show the limits obtained using the
calculation of Ref.~\cite{BAURBERGER} where the NLO corrections are
approximated by a constant $k$-factor given by
$k=1+8\alpha_s/9\pi\approx 1.34$. This calculation has been used to
extract bounds on $Z\gamma V$ couplings from $Z(\to\bar\nu\nu)\gamma$
production at the Tevatron~\cite{d0nndata}. Since a constant $k$-factor
overestimates the cross section for the jet veto cut imposed in the
current experimental analysis of the $Z(\to\bar\nu\nu)\gamma$ channel, 
the sensitivity limits obtained are $10-13\%$
better than those found using the full NLO 0-jet calculation,
depending on the integrated luminosity. Results similar to those shown
in Fig.~\ref{FIG:FOURTEEN} and Table~\ref{TAB:TWO} are obtained for
$h_{3,4}^\gamma$.

The 95\% CL limit contours in the $h_{30}^Z - h_{40}^Z$ plane for
$Z(\to\ell^+\ell^-)\gamma$ production at the LHC are shown in 
Fig.~\ref{FIG:FIFTEEN}, assuming an integrated luminosity of
100~fb$^{-1}$. Table~\ref{TAB:THREE} summarizes the LO and NLO
sensitivity bounds for $pp\to Z\gamma+X\to\ell^+\ell^-\gamma+X$ and
$pp\to Z\gamma+X\to p\llap/_T\gamma+X$ at $\sqrt{s}=14$~TeV with
$\int\!{\cal L}dt=10~{\rm fb}^{-1}$ and 100~fb$^{-1}$. 
At LHC energies, the inclusive ${\cal O}(\alpha_s)$ QCD
corrections in the SM considerably change the shape of the $p_T(\gamma)$
distribution (see Fig.~\ref{FIG:ACZEROJ}b). As a result, the inclusive NLO QCD
corrections reduce the sensitivity to anomalous couplings by 7~--~10\%.
As the integrated luminosity increases, larger transverse momenta become
accessible. The difference between the LO and NLO sensitivity bounds for
100~fb$^{-1}$ therefore is slightly larger than for 10~fb$^{-1}$. In
Sec.~IIID we have demonstrated that the size of the ${\cal O}(\alpha_s)$
QCD corrections at the LHC in the high $p_T(\gamma)$ region can be
reduced by vetoing hard jets in the central rapidity region. The
sensitivity bounds obtained for the $Z\gamma+0$~jet channel are about
5\% better than those found for the inclusive NLO case. However, because
the NLO 0-jet cross section is smaller than the LO cross section for the
jet definition we use (see 
Fig.~\ref{FIG:ACZEROJ}b), the limits obtained in the NLO $Z\gamma+0$~jet
case are slightly worse than those extracted from the LO cross section. 

As we have mentioned in Sec.~IIID, the NLO $Z\gamma+0$~jet differential 
cross section is more stable to variations of the factorization scale
$Q^2$ than the LO and inclusive NLO $Z\gamma+X$ cross sections. The
systematic errors which originate from the choice of $Q^2$ will thus be
smaller for bounds derived from the NLO $Z\gamma+0$~jet than those
obtained from the inclusive NLO $Z\gamma+X$ or the LO cross section.

As discussed in Ref.~\cite{BAURBERGER}, the limits which can be achieved
are sensitive to the form and the scale of the form factor. For example,
doubling the form factor scale to $\Lambda_{FF}=1$~TeV at the Tevatron
improves the bounds by almost a factor~3. The dependence of the limits
on $\Lambda_{FF}$ can be understood easily from Figs.~\ref{FIG:ACTEV}
and~\ref{FIG:ACLHC}. The improvement in sensitivity with increasing 
$\Lambda_{FF}$ is due to the additional events at large $p_T(\gamma)$
which are suppressed by the form factor if the scale $\Lambda_{FF}$ has
a smaller value. To a lesser degree, the bounds also depend on the power
$n$ in the form factor. Reducing $n$ allows for additional high
$p_T(\gamma)$ events and therefore leads to a somewhat increased
sensitivity to the low energy values of the anomalous couplings. It
should be noted, however, that $n$ must be larger than 3/2 (5/2) for
$h^V_{1,3}$ ($h^V_{2,4}$) in order to preserve $S$-matrix
unitarity~\cite{BAURBERGER}.

The bounds derived in this section are quite conservative. Using more
powerful statistical tools than the simple $\chi^2$ test we performed
can lead to considerably improved limits~\cite{GREG}. The effect of the
NLO QCD corrections on the sensitivity bounds, however, does not depend
on the technique used to extract them.

\section{SUMMARY}

Hadronic $Z\gamma$ production can be used to probe for non-standard
self interactions of the photon and $Z$ boson. The experimental limits 
for non-standard $ZZ\gamma$ and $Z\gamma\gamma$ 
couplings~\cite{d0dib,cdflldata,d0lldata,d0nndata} so far have been
based on leading order calculations~\cite{BAURBERGER}. 
In this paper we have presented an ${\cal O}(\alpha_s)$ calculation of 
the reactions $p\,p\hskip-7pt\hbox{$^{^{(\!-\!)}}$} \rightarrow Z 
\gamma + X \rightarrow \ell^+ \ell^- \gamma + X$ and 
$p\,p\hskip-7pt\hbox{$^{^{(\!-\!)}}$} \rightarrow Z \gamma + X
\rightarrow p\llap/_T\gamma + X$ for general $ZZ\gamma$ and
$Z \gamma \gamma$ couplings. The leptonic decay of the $Z$ boson has
been included in the narrow width approximation in our calculation.
Decay spin correlations are correctly taken into account, except in the 
finite virtual contribution. The finite
virtual correction term contributes only at the few per cent level to
the total NLO cross section, thus decay spin correlations can be safely
ignored here. Photon radiation from the final state lepton line is not
taken into account; effects
from $Z\to\ell^+\ell^-\gamma$ decays can easily be suppressed by
imposing a $\ell\ell\gamma$ invariant mass cut and a cut on the lepton
photon separation.

The $p_T(\gamma)$ differential cross section is very sensitive to
nonstandard $Z\gamma V$ ($V=Z,\,\gamma$) couplings. QCD corrections
change the shape of this distribution. The shape change is due to a
logarithmic enhancement factor in the $qg$ and $\bar qg$ real emission
subprocesses which appears in the high $p_T(\gamma)$ region of phase
space where the photon is balanced by a high $p_T$ quark which
radiates a soft $Z$ boson. The logarithmic enhancement factor combined
with the large gluon density at high center of mass energies make the 
${\cal O}(\alpha_s)$ corrections quite large for $p_T(\gamma)\gg m_Z$. 
Since the Feynman diagrams responsible for the enhancement at large 
$p_T(\gamma)$ do not involve any $Z\gamma V$ couplings, inclusive 
${\cal O}(\alpha_s)$ QCD corrections to $Z\gamma$ production tend to 
reduce the sensitivity to anomalous couplings. 

At the Tevatron, $Z\gamma$ production proceeds mainly via $q \bar q$
annihilation. Here the main effect of the QCD corrections is an increase
of the cross section by about 20~--~25\%. The sensitivity limits derived
from the inclusive NLO $Z\gamma + X$ cross section at the Tevatron are
thus up to 10\% better than those obtained from the LO cross section. If
a jet veto is imposed, the limits are almost identical to those
found using the LO calculation. In its $p\bar p\to Z\gamma\to 
p\llap/_T\gamma$ analysis, the D\O\ Collaboration imposes a $p_T(j)
<15$~GeV 
requirement, and uses the calculation of Ref.~\cite{BAURBERGER} where the
effect of the NLO QCD corrections is approximated by a simple $k$-factor
to extract sensitivity limits. We found that the bounds obtained from
the full NLO
$Z(\to\bar\nu\nu)\gamma+0$~jet calculation are $10-13\%$ weaker than 
those derived using the calculation used in the experimental analysis.

At the LHC, $qg$ fusion significantly contributes to $Z\gamma$
production and the change in the slope of the $p_T(\gamma)$ distribution
caused by the NLO QCD corrections is quite pronounced. As a result, the
limits on $ZZ\gamma$ and $Z\gamma\gamma$ couplings extracted from the
inclusive NLO $Z\gamma+X$ cross section are up to 10\% weaker than those
extracted using the LO calculation. The size of the QCD corrections at
large photon transverse momenta can be reduced considerably, and a
fraction of the sensitivity to $Z\gamma$ couplings which was lost at
the LHC may be regained by imposing a jet veto. The improvement,
however, is moderate. 

Although a jet veto does not have a large effect on the sensitivity
bounds at the Tevatron or the LHC, extracting limits from the $Z\gamma+0$~jet
channel has the advantage of a reduced uncertainty from the variation of
the factorization scale $Q^2$; the dependence of the NLO $Z\gamma+0$~jet
cross section on $Q^2$ is significantly smaller than 
that of the inclusive NLO and the LO $Z\gamma$ cross section.

The effect of QCD corrections on the sensitivity limits for anomalous
gauge boson couplings in $Z\gamma$ production is significantly smaller
than for $W\gamma$, $WZ$, and $W^+W^-$ production. In $W\gamma$ and $WZ$ 
production, the SM Born cross section is suppressed due to the 
appearance of an exact or approximate radiation 
zero~\cite{RADZERO,BHOR}. In $W^+W^-$ production, the strong
correlation of the $W$ helicities in the SM, together with the effect of
kinematic cuts, is responsible for the larger effect of the QCD
corrections~\cite{BHO2}. 

%
\acknowledgements

We would like to thank S.~Errede, B.~Harris, and G.~Landsberg for useful 
and stimulating discussions. Two of us (U.B. and T.H.) are grateful to 
the Fermilab Theory Group,
where part of this work was carried out, for its generous hospitality.
This work has been supported in part by Department of Energy 
grant No.~DE-FG03-91ER40674, NSF grant PHY-9600770 and the Davis
Institute for High Energy Physics.

%
%

%
\newpage
%
\widetext
\begin{table}
\caption{Sensitivities achievable at the 95\% confidence
level for the anomalous $Z\gamma V$ couplings $h_{30}^V$ and
$h_{40}^V$ ($V=Z,\,\gamma$) in $\protect{p\bar p\rightarrow Z\gamma + 
X\rightarrow\ell^+ \ell^-\gamma +X}$, $\ell=e,\,\mu$, at the Tevatron 
($\protect{\sqrt{s}=1.8}$~TeV) with $\int\!{\cal L}dt=1$~fb$^{-1}$. 
The limits for each coupling apply for arbitrary values of the other 
couplings listed in this table. 
The $CP$ violating couplings $h^V_{1,2}$ are assumed to take their SM values.
For the form factor we use the form of Eq.~(\protect{\ref{EQ:DIPOLES}})
with $n=3$ ($n=4$) for $h_3^V$ ($h_4^V$) and $\Lambda_{FF}=0.5$~TeV.
The cuts summarized in Sec.~IIIB are imposed. For the jet definition,
we have used Eq.~(\protect{\ref{EQ:TEVJET}}). \protect{\\} }
\label{TAB:ONE}
\begin{tabular}{cccc}
 & & & \\ [-7.mm]
\multicolumn{1}{c}{coupling}
&\multicolumn{1}{c}{LO}
&\multicolumn{1}{c}{NLO incl.}
&\multicolumn{1}{c}{NLO 0-jet}\\
\tableline
 & & & \\ [-6.mm]
$|h_{30}^Z|$ & 0.62 & 0.58 & 0.61 \\
$|h_{40}^Z|$ & 0.136 & 0.124 & 0.130 \\
\tableline
 & & & \\ [-6.mm]
$|h_{30}^\gamma|$ & 0.65 & 0.64 & 0.68 \\
$|h_{40}^\gamma|$ & 0.141 & 0.138 & 0.148 \\
\end{tabular}
\end{table}
\newpage
\begin{table}
\caption{Sensitivities achievable at the 95\% confidence
level for the anomalous $ZZ\gamma$ couplings $h_{30}^Z$ and
$h_{40}^Z$ in $\protect{p\bar p\rightarrow Z\gamma + 
X\rightarrow p\llap/_T\gamma +X}$ at the Tevatron 
($\protect{\sqrt{s}=1.8}$~TeV) for a) $\int\!{\cal L}dt=1~{\rm
fb}^{-1}$, and b) $\int\!{\cal L}dt=10~{\rm fb}^{-1}$. Shown are the
limits obtained from the LO calculation, the full NLO 0-jet
differential cross section, and the calculation of 
Ref.~[\protect{\ref{BABE}}]. The limits for 
each coupling apply for arbitrary values of the other coupling listed in 
this table. The $CP$ violating couplings $h^Z_{1,2}$ and all 
$Z\gamma\gamma$ couplings are assumed to vanish.
For the form factor we use the form of Eq.~(\protect{\ref{EQ:DIPOLES}})
with $n=3$ ($n=4$) for $h_3^Z$ ($h_4^Z$) and $\Lambda_{FF}=0.5$~TeV.
The cuts summarized in Sec.~IIIB are imposed. 
For the jet definition,
we have used Eq.~(\protect{\ref{EQ:TEVJET}}). \protect{\\} }
\label{TAB:TWO}
\begin{tabular}{cccc}
 & & & \\ [-7.mm]
\multicolumn{4}{c}{a) $\int\!{\cal L}dt=1~{\rm fb}^{-1}$} \\
\multicolumn{1}{c}{coupling}
&\multicolumn{1}{c}{LO}
&\multicolumn{1}{c}{NLO 0-jet}
&\multicolumn{1}{c}{NLO appr.}
\\
\tableline
 & & & \\ [-6.mm]
$|h_{30}^Z|$ & 0.55 & 0.53 & 0.46 \\
$|h_{40}^Z|$ & 0.108 & 0.104 & 0.091 \\
\tableline
\tableline
 & & & \\ [-7.mm]
\multicolumn{4}{c}{b) $\int\!{\cal L}dt=10~{\rm fb}^{-1}$} \\
\multicolumn{1}{c}{coupling}
&\multicolumn{1}{c}{LO}
&\multicolumn{1}{c}{NLO 0-jet}
&\multicolumn{1}{c}{NLO appr.}
\\
\tableline
 & & & \\ [-6.mm]
$|h_{30}^Z|$ & 0.30 & 0.29 & 0.26 \\
$|h_{40}^Z|$ & 0.055 & 0.053 & 0.047 \\
\end{tabular}
\end{table}
\newpage
\begin{table}
\caption{Sensitivities achievable at the 95\% confidence
level for the anomalous $ZZ\gamma$ couplings $h_{30}^Z$ and
$h_{40}^Z$ in a) $\protect{pp\rightarrow Z\gamma +
X\to\ell^+\ell^-\gamma+X}$ and b) $\protect{pp\rightarrow Z\gamma + 
X\rightarrow p\llap/_T\gamma +X}$, at the LHC
($\protect{\sqrt{s}=14}$~TeV). Results are shown for integrated
luminosities of 10~fb$^{-1}$ and 100~fb$^{-1}$. The limits for each
coupling apply for arbitrary values of the other coupling listed in 
this table. The $CP$ violating couplings $h^Z_{1,2}$ and all 
$Z\gamma\gamma$ couplings are assumed to vanish.
For the form factor we use the form of Eq.~(\protect{\ref{EQ:DIPOLES}})
with $n=3$ ($n=4$) for $h_3^Z$ ($h_4^Z$) and $\Lambda_{FF}=3$~TeV.
The cuts summarized in Sec.~IIIB are imposed. For the jet definition,
we have used Eq.~(\protect{\ref{EQ:LHCJET}}). \protect{\\} }
\label{TAB:THREE}
\begin{tabular}{cccc}
 & & & \\ [-7.mm]
\multicolumn{4}{c}{a) $pp\rightarrow Z\gamma+X\to\ell^+\ell^-\gamma+X$} \\
\tableline
\multicolumn{4}{c}{ $\int\!{\cal L}dt=10~{\rm fb}^{-1}$} \\
\multicolumn{1}{c}{coupling}
&\multicolumn{1}{c}{LO}
&\multicolumn{1}{c}{NLO incl.}
&\multicolumn{1}{c}{NLO 0-jet}
\\
\tableline
$|h_{30}^Z|$ & $4.6\times 10^{-3}$ & $5.0\times 10^{-3}$ & $4.7\times
10^{-3}$  \\
$|h_{40}^Z|$ & $3.6\times 10^{-5}$  & $3.9\times 10^{-5}$ & $3.7\times 
10^{-5}$ \\
\tableline
 & & & \\ [-7.mm]
\multicolumn{4}{c} {$\int\!{\cal L}dt=100~{\rm fb}^{-1}$} \\
\multicolumn{1}{c}{coupling}
&\multicolumn{1}{c}{LO}
&\multicolumn{1}{c}{NLO incl.}
&\multicolumn{1}{c}{NLO 0-jet}
\\
\tableline
$|h_{30}^Z|$ & $2.5\times 10^{-3}$ & $2.8\times 10^{-3}$ & $2.6\times
10^{-3}$  \\
$|h_{40}^Z|$ & $1.7\times 10^{-5}$  & $1.9\times 10^{-5}$ & $1.8\times 
10^{-5}$ \\
\tableline
\tableline
\multicolumn{4}{c}{b) $pp\rightarrow Z\gamma+X\to p\llap/_T\gamma+X$} \\
\tableline
\multicolumn{4}{c}{ $\int\!{\cal L}dt=10~{\rm fb}^{-1}$} \\
\multicolumn{1}{c}{coupling}
&\multicolumn{1}{c}{LO}
&\multicolumn{1}{c}{NLO incl.}
&\multicolumn{1}{c}{NLO 0-jet}
\\
\tableline
$|h_{30}^Z|$ & $3.4\times 10^{-3}$ & $3.7\times 10^{-3}$ & $3.5\times
10^{-3}$  \\
$|h_{40}^Z|$ & $2.5\times 10^{-5}$  & $2.7\times 10^{-5}$ & $2.6\times 
10^{-5}$ \\
\tableline
 & & & \\ [-7.mm]
\multicolumn{4}{c} {$\int\!{\cal L}dt=100~{\rm fb}^{-1}$} \\
\multicolumn{1}{c}{coupling}
&\multicolumn{1}{c}{LO}
&\multicolumn{1}{c}{NLO incl.}
&\multicolumn{1}{c}{NLO 0-jet}
\\
\tableline
$|h_{30}^Z|$ & $1.9\times 10^{-3}$ & $2.0\times 10^{-3}$ & $1.9\times
10^{-3}$  \\
$|h_{40}^Z|$ & $1.2\times 10^{-5}$  & $1.3\times 10^{-5}$ & $1.2\times 
10^{-5}$ \\
\end{tabular}
\end{table}
\newpage
%
%
\newpage

\newpage

\newpage

\newpage

\newpage

\newpage

\newpage

\newpage

\newpage
%
\newpage
%
\newpage
%
\newpage
%
\newpage
%
\newpage
%
%
\newpage
%
\end{document}